\documentclass[prl,aps,amssym,superscriptaddress,nofootinbib,floatfix,reprint,notitlepage]{revtex4-2} 

\usepackage{xcolor}
\usepackage{amsmath,amssymb,bbold,bm}
\usepackage{mleftright}
\usepackage{graphicx}
\usepackage[normalem]{ulem}
\usepackage{mleftright}
\usepackage{cancel}
\usepackage{comment}
\usepackage{blindtext}
\usepackage{titlesec}
\usepackage{mathdots}
\usepackage{mathtools}
\usepackage{caption}
\usepackage{currfile}

\usepackage[pdfpagelabels,colorlinks=true,breaklinks=true,linktocpage=true,linkcolor=red,urlcolor=blue,citecolor=blue]{hyperref}

\def\dul#1{{\underline{#1}}}

\newcommand{\rr}{\hspace{-.075cm}}
\newcommand{\sbraket}[1]{\langle #1 \rangle}
\newcommand{\be}{\begin{equation}}
\newcommand{\ben}{\begin{equation*}}
\newcommand{\ee}{\end{equation}}
\newcommand{\een}{\end{equation*}}
\newcommand{\bs}{\begin{split}}
\newcommand{\es}{\end{split}}
\newcommand{\bmx}{\begin{array}}
\newcommand{\emx}{\end{array}}

\newcommand{\bea}{\begin{eqnarray}}
\newcommand{\bean}{\begin{eqnarray*}}
\newcommand{\eea}{\end{eqnarray}}
\newcommand{\eean}{\end{eqnarray*}}
\newcommand{\dg}{^{\dagger}}
\newcommand{\dn}{^{\vphantom{\dagger}}}

\newcommand{\lr}{\leftrightarrow}

\newcommand{\ua}{\uparrow}
\newcommand{\da}{\downarrow}
\newcommand{\Ua}{\Uparrow}
\newcommand{\Da}{\Downarrow}
\newcommand{\bb}[1]{\mathbb{#1}}
\newcommand{\qqquad}{\qquad\qquad\qquad}
\newcommand{\so}{\qquad\rightarrow\qquad}
\newcommand{\So}{\qquad\Rightarrow\qquad}

\newcommand{\andd}{\qquad\text{and}\qquad}

\newcommand{\eps}{\epsilon}

\newcommand{\tpsi}{\tilde{\psi}}

\newcommand{\pref}[1]{(\ref{#1})}

\newcommand{\abs}[1]{\left\vert #1 \right\vert}
\newcommand{\bra}[1]{\left\langle #1 \right\vert}
\newcommand{\ket}[1]{\left\vert #1\right\rangle}

\newcommand{\braaa}[1]{\left\langle\hspace{-.07cm}\left\langle\hspace{-.07cm}\left\langle #1 \right\vert\right.\right.}
\newcommand{\kett}[1]{\left.\left\vert #1\right\rangle\hspace{-.07cm}\right\rangle}
\newcommand{\kettt}[1]{\left.\left.\left\vert #1\right\rangle\hspace{-.07cm}\right\rangle\hspace{-.07cm}\right\rangle}
\newcommand{\braket}[1]{\left\langle #1\right\rangle}

\newcommand{\mat}[1]{\left(\bmx{cc}#1\emx\right)}
\newcommand{\matc}[2]{\left(\bmx{#1}#2\emx\right)}

\newcommand{\matn}[1]{\bmx{cc}#1\emx}
\newcommand{\matl}[1]{\bmx{ll}#1\emx}

\newcommand{\bw}[1]{\begin{widetext}}
\newcommand{\ew}[1]{\end{widetext}}

\newcommand{\red}[1]{{\color{red} #1}}
\newcommand{\gray}[1]{}

\newcommand{\blue}[1]{{#1}}

\newcommand{\nothing}[1]{}

\begin{document}

\title{Coherent manipulation of Kondo Majoranas in two-channel Kondo setups}
\author{Yashar Komijani\,$^{*}$}
 \affiliation{ Department of Physics, University of Cincinnati, Cincinnati, Ohio, 45221, USA}
\author{C.~J.~Bolech}
 \affiliation{ Department of Physics, University of Cincinnati, Cincinnati, Ohio, 45221, USA}
 \affiliation{Instituto de Ciencia de Materiales de Madrid (ICMM), Consejo Superior de Investigaciones Científicas (CSIC), Sor Juana Inés de la Cruz 3, E-28049 Madrid, Spain}
\affiliation{Departamento de Física Teórica de la Materia Condensada, Condensed Matter Physics Center (IFIMAC), Universidad Autónoma de Madrid, E-28049 Madrid, Spain}
\date{\today}
\begin{abstract}
We study coherent manipulation of Majorana zero modes emerging in overscreened two-channel Kondo systems. Using compactified lattice models, we show that these interacting Kondo Majoranas support non-local qubits and admit teleportation, fusion, and braiding operations. In particular, we identify a distinction between non-topological and genuinely topological Y-junction geometries, the latter realizing a non-Abelian geometric holonomy. Our results establish a proof-of-principle route toward coherent control of non-Abelian anyons beyond conventional free-fermion platforms.
\end{abstract}
\maketitle

\emph{\blue{Introduction}} --
Multi-channel Kondo systems provide one of the simplest settings in which strong correlations give rise to emergent non-Fermi liquid behavior and fractionalized boundary degrees of freedom \cite{Nozieres80,Andrei84,Affleck92,Emery1993,Affleck1993,Bolech2002}. In particular, two-channel Kondo (2CK) impurities have recently been argued to host localized Majorana zero modes (MZMs) embedded within the many-body Kondo cloud \cite{Lopes2020,Komijaniqubit}. Unlike the MZMs arising in non-interacting topological superconductor models \cite{Kitaev2001,Lutchyn2010,Oreg2010,Microsoft2025}, these excitations emerge intrinsically from strong correlations and carry an emergent gauge structure. 

Recent advances in charge-Kondo devices based on single-electron transistors (SET) coupled to integer quantum Hall (IQH) edge states
\cite{Iftikhar2015,Iftikhar2018,Pouse2023}, including recent measurements of residual impurity entropy \cite{Piquard2026}, have renewed interest in whether emergent Kondo anyons can be coherently manipulated and utilized for topological quantum information processing. Previous works have proposed non-local qubit architectures and measurement-based protocols involving Kondo Majoranas \cite{Gobay2022,Lotem2022,Lotem2023,Ren2024,Sen2024}. However, a transparent framework for coherent manipulation and braiding of these interacting anyons has remained elusive.

In this work, we develop a conceptual architecture for manipulating Kondo Majorana modes within compactified lattice representations of the 2CK fixed point. The compactified description \cite{Ljepoja2024A,Ljepoja2024B,Ljepoja2024C} maps the finite-coupling 2CK problem onto an effectively strong-coupling framework, rendering the emergent non-local Majorana degrees of freedom operationally accessible within finite-dimensional Hilbert spaces. This representation allows us to explicitly construct and manipulate Kondo Majorana qubits, including coherent rotations, fusion protocols, teleportation, and braiding operations.

A central result of this work is that the adiabatic transport of Kondo Majoranas admits a simple geometric interpretation in terms of parallel transport within a degenerate ground-state (GS) manifold. In this formulation, braiding emerges naturally as a non-Abelian Berry holonomy associated with adiabatic evolution through Y-junction geometries. The resulting framework provides a concrete route toward coherent control of interacting anyons and clarifies the geometric structure hidden within multi-channel Kondo physics.

\begin{figure}[tp]
\includegraphics[width=1\linewidth]{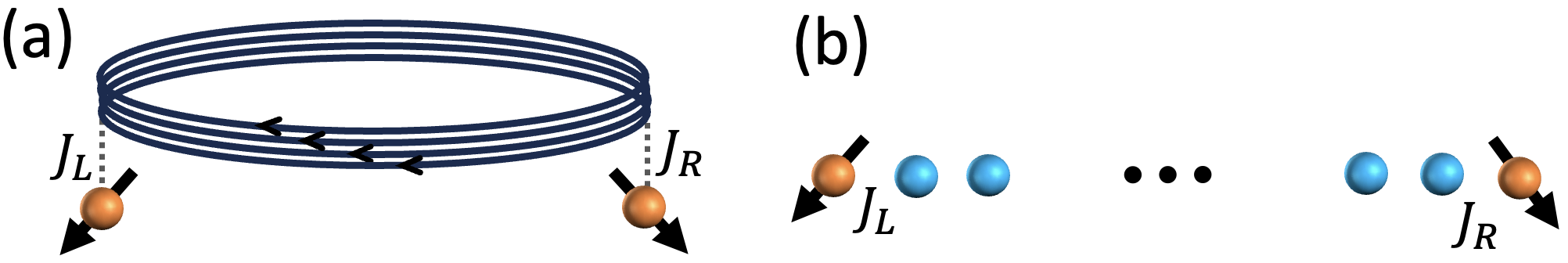}
\caption{\small\raggedright (a) A two-impurity two-channel Kondo (2CK) setup using 1D edge states. (b) Compactified tight-binding chain.\vspace{-.5cm}}\label{Fig1}
\end{figure}

\emph{\blue{Original and effective models}} -- 
Following \cite{Lopes2020,Gobay2022,Lotem2022,Lotem2023,Ren2024,Sen2024} we consider multi-impurity 2CK systems composed of local moments coupled to a chiral edge state. A representative two-impurity setup [Fig.\,\ref{Fig1}(a)]. The edge state consists of two-channel spinful chiral fermions interacting with spin-$1/2$ local moments through the Hamiltonian \cite{Ftnote2}
\vspace{-.3cm}
\be
H=H_0+\sum_{j=1}^N\vec S_j\cdot 
\sum_{a=1}^2J_K^{(a)}\psi\dg_{a\alpha}(x_j)[\vec{\dul\sigma}]_{\alpha\beta}\psi\dn_{a\beta}(x_j),
\vspace{-.2cm}
\ee
where $\psi_{a\alpha}$ has spin $\alpha,\beta=\ua,\da$ (with implicit summation) and flavor (or channel) $a=1,2$ indices. 
A key feature of this model is that only the spin sector of the conduction electrons participates in the Kondo interaction, while the charge in each channel remains conserved. Consequently, the charge and flavor sectors decouple from the impurity dynamics, leaving only spin and spin-flavor degrees of freedom coupled to the impurity \cite{SM}. This is particularly transparent in the bosonized formulation and leads to a substantially reduced and computationally economical Hilbert space.

Refermionizing the spin and spin-flavor sectors (without invoking the Emery--Kivelson transformation), the problem reduces to the compactified 2CK model \cite{Coleman1995,Bulla1997b,Bulla1997}. Grouping the resulting fermions into a Nambu spinor $\tilde\Psi$, the interaction takes the form $\vec S\cdot \tilde\Psi\dg(J^{(1)}_K\vec{\dul\tau}+J^{(2)}_K\vec{\dul\sigma})\tilde\Psi$, where $\vec{\dul\sigma}$ and $\vec{\dul\tau}$ act on (pseudo)spin and particle-hole isospin degrees of freedom, respectively. Finally, the compactified model admits an efficient tight-binding representation [Fig.\,\ref{Fig1}(b)], enabling exact diagonalization (ED) of finite systems and real-time evolution through the quantum Liouville equation $i\hbar\dot\rho=[H(t),\rho]$.

\emph{\blue{Single 2CK impurity analysis}} -- 
With a single impurity, the problem can be mapped \cite{SM} to a tight-binding chain of spinful electrons end-coupled to an impurity spin
\vspace{-.5cm}
\be
H=-\sum_{n=1}^\infty(itc\dg_nc\dn_{n+1}+h.c.)+\vec S\cdot C_1\dg(J_K^{(1)}\vec{\dul\tau}+J_K^{(2)}\vec{\dul\sigma})C_1,
\vspace{-.3cm}
\ee
where $c$ is spinful and $C\dg\equiv (c\dg\,\,c^T)$ contains both Nambu sectors of $c$. We have included a factor of $i$ in the tunneling amplitude for later convenience. The significance of compactification is that the (pseudo)spin and isospin of the fermions have different parities and cannot be simultaneously non-zero \cite{Ftnote4}. Therefore, the impurity spin is not overscreened, and the strong Kondo coupling $J_K=\infty$ is the stable fixed point in this compactified model \cite{Coleman95e}. Consequently, a Kondo length scale $\xi_K=v_F/T_K$ remains well defined in the compactified model, even though in the original formulation the corresponding correlations are encoded nonlocally and decay algebraically with distance.

\emph{\blue{Symmetries}} -- The problem has conduction-electron fermion-parity symmetry
$(-1)^F=\prod_j\sqrt{2i}\chi_j$. In the presence of Kondo coupling, neither the conduction-electron spin, the isospin, nor their sum $\vec{\cal J}=\sum_n C_n^\dagger(\vec{\dul\sigma}+\vec{\dul\tau})C_n$ is conserved. However, the total charge
$\vec{\cal I}=\vec{\cal J}+\vec S$ remains conserved.

For $J_K=J_K'$, there is a (parity) self-duality $\sigma\lr\tau$ which is reflected in the discrete Z$_2$ symmetry $\chi^0\to-\chi^0$, but this is broken in presence of a channel asymmetry.

\emph{\blue{Kondo Majorana}} -- 
A single impurity at the strong Kondo coupling limit has the following structure; 
Using four local MZMs $\{\chi^x,\chi^y,\chi^z,\chi^0\}$ to represent the spinful $c$-fermions (lattice representation of $\tilde\Psi$), by $c_\ua=-(\chi^x-i\chi^y)/\sqrt 2$ and $c_\da=(\chi^z-i\chi^0)/\sqrt{2}$ allows us  in the channel symmetric case to write the Kondo interaction as $H_{int}=J_KS^a{\cal J}^a$ where ${\cal J}^a=-\frac{i}2\eps^{abc}\chi^b\chi^c$ transform under the total SU(2). It can be easily checked that ${\cal J}^2=3/4$ and thus $\vec{\cal J}$ is a spin-1/2 \emph{fully screening} the impurity, in this compactified form. This local MZM representation contrasts with the original formulation, in which overscreening arises from an SU(2)$_2$ current constructed from delocalized chiral Majorana modes.

Note that only three Majoranas ($x,y,z$ with the gauge we have chosen)  appear in the interaction and $\chi^0$ is decoupled. Equivalently, ${\cal J}$ with three MZMs has an anomalous Hilbert space, whereas with four MZMs, it has two representations with opposite fermion parities.

\begin{figure}[tp!]
\includegraphics[width=\linewidth]{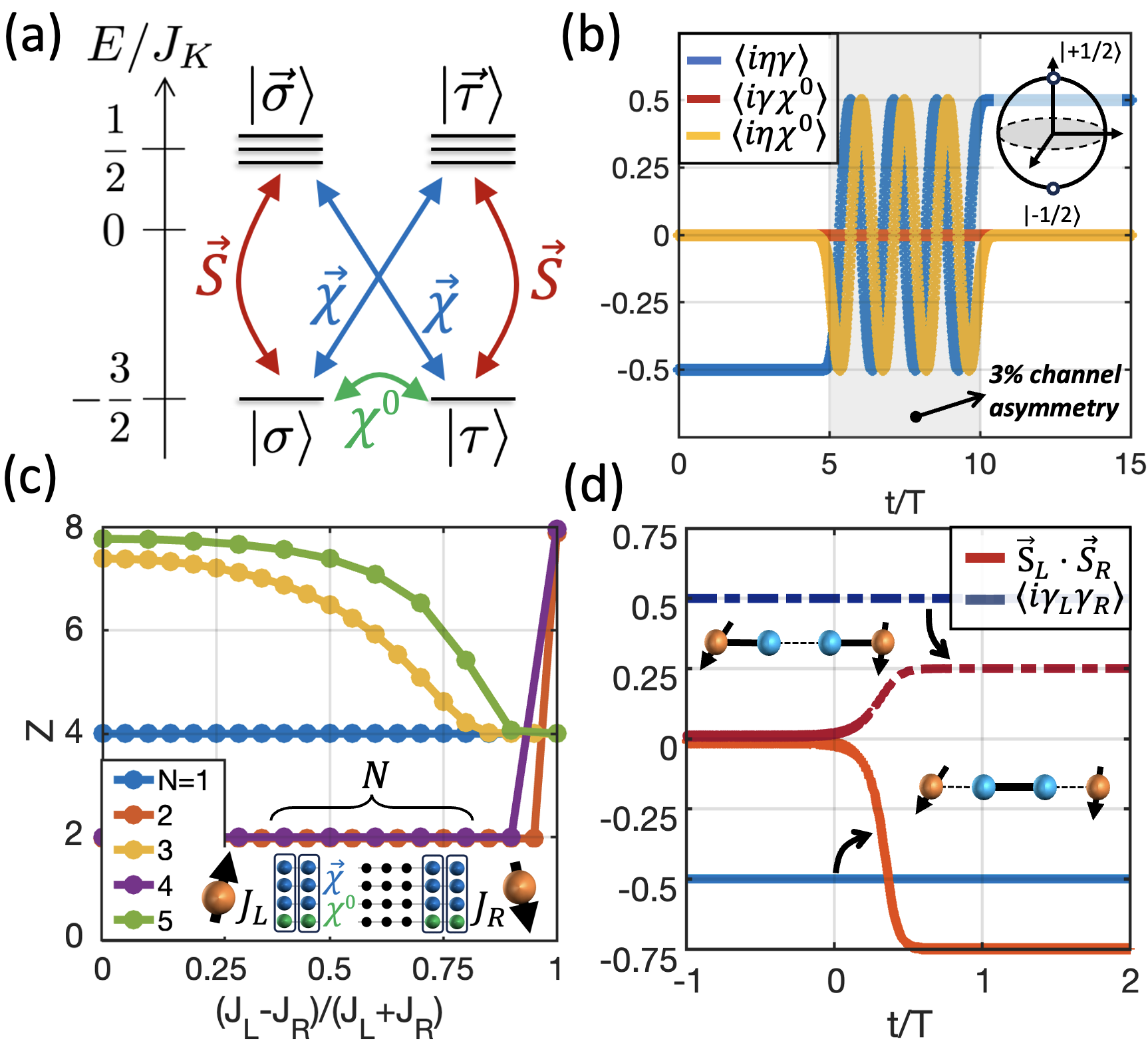}\vspace{-.2cm}
\caption{\small\raggedright (a) Energy levels of the single-site model. Various transitions are highlighted. (b) Rabi oscillations of the non-local qubit between $\gamma$ and the ancillary $\eta$ driven by channel anisotropy. (c) The GSD of two 2CK impurities connected by a chain of $N$ sites in the $J_K\gg t$ limit, represented by the partition function $Z$. (d) Kondo Majorana fusion onto the total spin of spin-singlet ($\vec S_1\cdot\vec S_2=-3/4$) or the spin-triplet ($\vec S_1\cdot\vec S_2=1/4$) depending on qubit state, $i\gamma_L\gamma_R=\mp 1/2$.\hspace{-.02cm}\vspace{-.5cm}}\label{Fig2}
\end{figure}

The structure of 8-dimensional Hilbert space, shown in Fig.\,\ref{Fig2}(a), consists the usual singlet $\ket{\sigma}$ and triplet $\vert{\vec\sigma}\rangle$ between the impurity spin and the electron pseudospin 
\be
\ket{\sigma}=\frac{\ket{\Ua\da-\Da\ua}}{\sqrt{2}},\quad \ket{\vec\sigma}=\big\{\ket{\Ua\ua},\frac{\ket{\Ua\da+\Da\ua}}{\sqrt 2},\ket{\Da\da}\big\},
\ee
and an additional copy of singlet $\ket{\tau}$ and triplet $\vert\vec\tau\rangle$ between the impurity spin and the electron isospin
\be
\ket{\tau}=\frac{\ket{\Ua\rr0-\Da\rr2}}{\sqrt{2}},\quad \ket{\vec\tau}=\big\{\ket{\Ua\rr2},\frac{\ket{\Ua\rr0+\Da\rr2}}{\sqrt 2},\ket{\Da\rr0}\big\}.
\ee
The GS is doubly degenerate with different fermionic parities. Beside the decoupled $\chi^0$, three new Majoranas fermions defined as $\gamma^a\equiv{2}\chi^aS^a$ commute with the interacting Hamiltonian $[\gamma^a,H_{int}]=0$, but they are not independent, and are equal when projected to the GS manifold. Therefore we define a single composite Majorana fermion $\gamma\equiv \frac{ 2}{3}\vec\chi\cdot\vec S$, normalized so that $\gamma^2=1/2$ on the GS sector.  In terms of $\gamma$ the interaction $H_{int}=-3J_K i\gamma\chi^0(-1)^F$ appears like a usual hybridization between Majorana fermions $\gamma$ and $\chi^0$, except for the nonlinearity introduced by the multiplying fermion parity factor $(-1)^F\equiv-4\chi^0\chi^x\chi^y\chi^z$, signifying that the two GSs have opposite fermionic parities. 

In the compactified model, the composite spin-fermion MZM $\gamma$ is localized near the impurity, spread over a length scale of Kondo cloud $2ta/T_K$, with lattice spacing $a$, whereas $\chi^0$ is a MZM delocalized throughout the bulk of the conduction electrons. However, in the original model $\gamma$ has a non-local admixture of the whole conduction band. The conduction $\chi^0$ and the localized $\gamma$ realize a delocalized charge qubit. The two act as Pauli matrices $\sigma^x$ and $-\sigma^y$ in the space of the GS, respectively, while their product $i\gamma\chi^0$ acts as $\sigma^z$.

\emph{\blue{Coupling to an auxiliary MZM}} -- 
It is instructive to couple the Kondo MZM to an auxiliary MZM defined by adding a fermionic state $\sqrt{2}f\dg=\eta'+i\eta''$ to the Hilbert space. Suppose we have the Hamiltonian $H=im\gamma\eta'$ with $0<m\ll J_K$.  The low energy sector has 4 states with opposite parities
\be
\ket{\pm}^e=\frac{\ket{\tau\,0_f}\mp i\ket{\sigma\,1_f}}{\sqrt{2}}, \quad \ket{\pm}^o=\frac{\ket{\sigma\,0_f}\pm i\ket{\tau\,1_f}}{\sqrt{2}}.
\ee
The energies are $E_\pm=\mp m/2$ for both even/odd sectors, within each again $\gamma$ acts as $\sigma^x$ and $\eta'$ acts as $-\sigma^y$. Hereafter, we are using a redundant $\sigma-\tau$ basis which relies on the decoupled $\chi^0$ MZM. In each case, this can be traded for a more dense description not involving $\chi^0$ \cite{SM}.

\emph{\blue{Rabi oscillations}} -- 
Given an ancillary MZM $\eta$ entangled with $\gamma$, the set of $(\gamma,\chi^0,\eta)$ forms a spin qubit (with a $Z_2$ gauge redundancy), which can be represented on a Bloch sphere. Initially, this effective spin points toward the north/south pole. Coupling between any two of the MZMs corresponds to a Zeeman field, which causes spin precession. In the experiment this can be naturally induced by channel anisotropy, modeled by the additional term $\Delta H=\Delta J_K\vec S\cdot C\dg(\vec\sigma-\vec\tau)C$ in the compactified Hamiltonian, equivalent to $\Delta H=\Delta J_K i\gamma\chi^0$ \cite{Sengupta94}, without the $(-1)^F$ factor. By adjusting the magnitude and duration of the pulse, an arbitrary rotation around the $i\gamma\chi^0$ axis can be achieved [Fig.\,\ref{Fig2}(b)]. On the other hand, keeping the asymmetry over a long period of time, polarizes the delocalized qubit in a specific direction by relaxation. Clearly, the Kondo MZM $\gamma$ holds its correlation with the ancillary MZM $\eta$ in the initial and final channel symmetric periods, and undergoes Rabi oscillations in the presence of channel asymmetry, providing a knob for robust manipulation of the non-local Kondo qubits.

\emph{\blue{Double 2CK impurity model}} -- As an example of higher complexity, we consider the problem of two 2CK impurities coupled to the same edge states, Fig.\,\ref{Fig1}(a). Using a convenient gauge $t\in \bb R$ \cite{Coleman1995}, the conduction band can be represented as a four-flavor chain of Majorana fermions $H_0=-it\sum_{n=1}^{N-1}\sum_{\mu=0}^3 \chi^\mu_n\chi^\mu_{n+1}$, whose length  $N$ is related to the length of the 1D state.The two-impurity 2CK problem \cite{Georges1997} can be mapped to the model
\vspace{-.2cm}
\be
H=H_0-\frac{i}2J_L\eps^{abc}S_L^a\chi_1^b\chi_1^c-\frac{i}2J_R\eps^{abc}S_R^a\chi_N^b\chi_N^c,\label{eq2c2CKchain}
\vspace{-.2cm}
\ee
as illustrated in the tight-binding model of Fig.\,\ref{Fig1}(b). 

\blue{\emph{The GS degeneracy} (GSD)} of the two-impurity 2CK model \pref{eq2c2CKchain} for small chains of $N=1\dots 5$ is shown in Fig.\,\ref{Fig2}(c) in the strong Kondo regime as a function of the left/right Kondo couplings' asymmetry using ED. The even and odd $N$ have markedly different behavior, and it can be traced back to periodic or anti-periodic boundary condition of the original $\psi$ fermions.

These results can be understood using two simple ideas; First, for odd $N$, each conduction Majorana chain has a zero-energy mode, contributing a factor of $\sqrt{2}$ to the GSD, whereas the GS is unique for even $N$.  Second, with $J_K/t\gg 1$, each spin detaches the nearby $\vec\chi$ sites, resulting in a MZM $\gamma=\frac{2}{3}\vec\chi\cdot\vec S$ and contributing a factor of $\sqrt{2}$ to GSD, but the delocalized $\chi^0$ chain remains uncut. This process, under $J_L=J_R$ symmetry shrinks the chain from both ends, maintaining the parity of $N\to N-2$, with a zero mode leftover for odd $N$.

The significance of this analysis is that it is independent of $N$ and can be extended to chains of arbitrary length, substantiating $\gamma_L$ and $\gamma_R$ as localized MZMs at the position of each impurity. For $N>1$ and large enough $J_K$, the two Kondo MZMs can coexist. The weak/strong Kondo coupling would determine if the Kondo MZMs have overlapping Kondo clouds. The latter can be used for fusion (even-$N$ case) and teleportation and braiding (odd-$N$ case) of Kondo MZMs.

The case of $N=1$ is special [Fig.\,\ref{Fig2}(c)]. In this case, the fermionic mode is overscreened by the impurities, resulting in a doublet under the conserved ${\cal I}$ charge. Additionally, writing the Hamiltonian as $H=-3i(J_L\gamma_L+J_R\gamma_R)\chi^0(-1)^F$, shows that $\gamma\propto J_L\gamma_L+J_R\gamma_R$ is coupled to the conduction electrons, contributing a factor of 2 to the GSD. This feature is shared with other odd chains in the $t\gg J_K$ limit, whereas even chains have decoupled spins with GSD of 4.

\begin{figure}[tp]
\includegraphics[width=\linewidth]{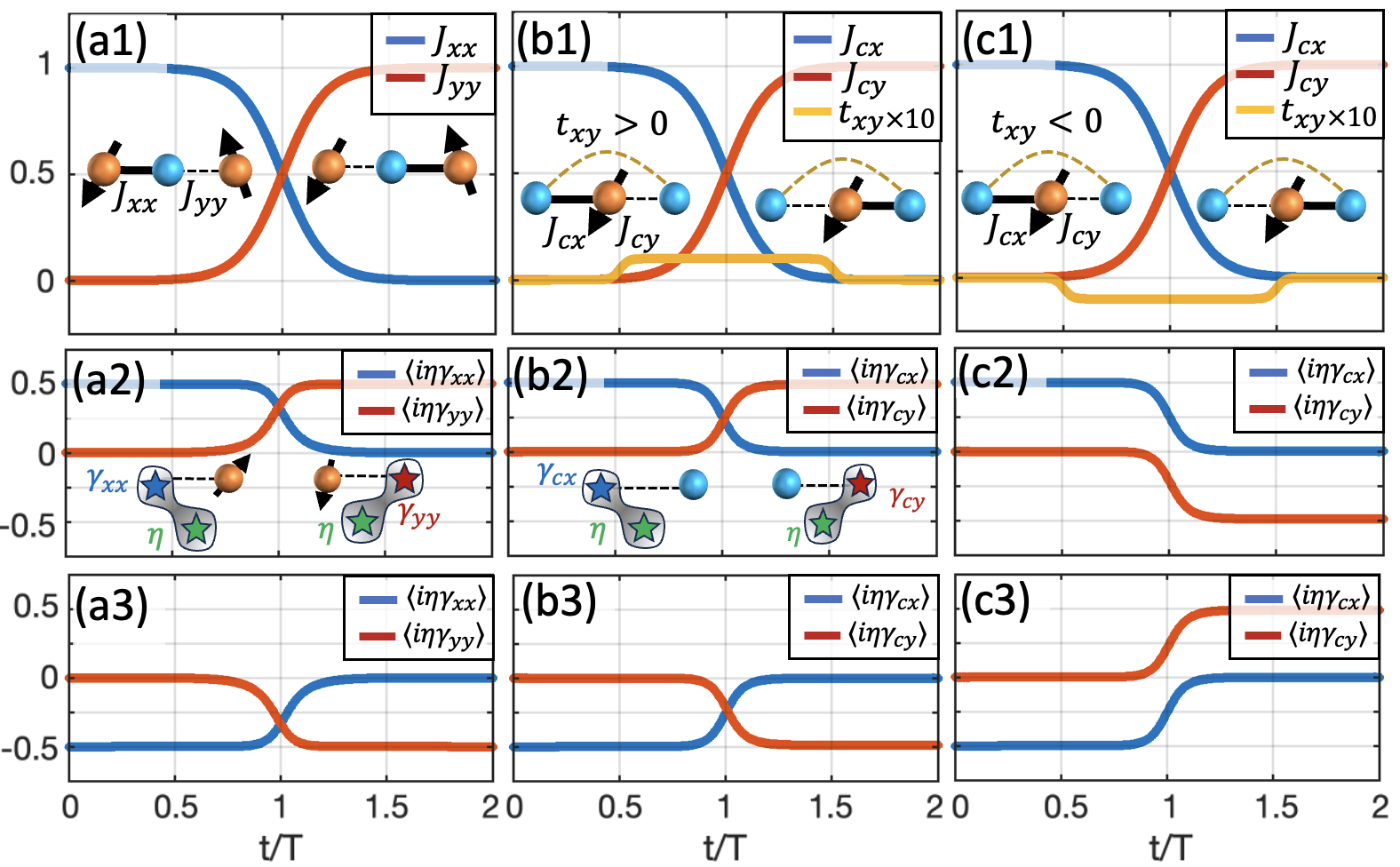}
\caption{\small\raggedright Deterministic and coherent MZM transfer, while preserving its entanglement with an ancillary MZM $\eta$,  
between two spins (a1-a3), and across a spin (b1-b3) and (c1-c3). Top row panels show the adiabatic variation of the Kondo and tunnel couplings, whereas the bottom two rows show the evolution of $\sbraket{i\eta\gamma_{xx}}$ or $\sbraket{i\eta\gamma_{cx}}$ (blue) and $\sbraket{i\eta\gamma_{yy}}$ or $\sbraket{i\eta\gamma_{cy}}$ (red) for two different initial states. Moving MZMs across spins requires a small tunneling $t_{xy}$, whose sign multiplies the transferred MZM entanglement (b and c).\vspace{-.5cm}
}\label{Fig3}
\end{figure}

\emph{\blue{Fusion}} -- 
The initialization and read-out of the Kondo MZMs can be achieved by transforming them into spins. This is achieved by gapping out zero modes in even chains. Starting from a $N=2$ chain (no zero-energy bulk mode) with large $J_K$ (small Kondo cloud), we increase $t/J_K$ to form significant overlap between $\gamma_L$ and $\gamma_R$. Since $S_L^a\sim i\gamma_L\chi_L^a$ and $S^a_R\sim i\gamma_R\chi_R^a$, and $\sbraket{i\vec\chi_L\cdot\vec\chi_R}=-3/2$ in the tunneling dominated regime, it is expected that $\sbraket{\vec S_1\cdot\vec S_2}$ is related to $\sbraket{i\gamma_L\gamma_R}$. 

Focusing on the even-parity sector, the two GSs are
\be
\ket{\psi}_\pm^e=\cos\alpha_\pm\ket{X}_\pm^e-\sin\alpha_\pm\ket{Y}_+^e,
\vspace{-.2cm}
\ee
where $\tan2\alpha_\pm=\sqrt{3}t/(\pm 2\bar J-t)$ and we have defined
\be
\hspace{-.35cm}\ket{X}_\pm^e=\frac{\ket{\sigma_L\sigma_R\pm i\tau_L\tau_R}}{\sqrt 2},\,\,
\ket{Y}_\pm^e=\frac{\ket{\vec\sigma_L\cdot\vec\sigma_R\mp i\vec\tau_L\cdot\vec\tau_R}}{\sqrt 6}.\hspace{-.2cm}
\vspace{-.1cm}
\ee
As $t/J$ increases the angles evolve as $2\alpha_+: 0\to 2\pi/3$ and $2\alpha_-: 0\to -\pi/3$. Using this, we find that the adiabatic evolution results in
\vspace{-.1cm}
\be
{}_+^e\sbraket{\psi\vert \vec S_1\cdot\vec S_2\vert\psi}_+^e\to -\frac{3}{4}, \qquad
{}_-^e\sbraket{\psi\vert \vec S_1\cdot\vec S_2\vert\psi}_-^e\to \frac{1}{4}.
\vspace{-.1cm}
\ee
Since ${\cal I}^z=0$, the triplet involved is $\ket{T_0}$. This enables not only a quantum non-demolition (QND) measurement of the Kondo qubit [Fig.\,\ref{Fig2}(d)], but also, when operated in reverse, initialization of its state.

\emph{\blue{Parity measurement}} -- 
With two Kondo qubits $(\gamma_1,\gamma_2)$ and $(\gamma_3,\gamma_4)$, after detaching from the conduction electrons, the spins fuse to distinct total spins. $\ket{S}_{12}\ket{S}_{34}$, $\{\ket{S}_{12}\ket{T_0}_{34}, \ket{T_0}_{12}\ket{S}_{34}\}$ and $\ket{T_0}_{12}\ket{T_0}_{34}$ fuse to $S_{\rm tot}=0$, $1$ and $2$, respectively. Measuring $\exp({i\pi S_{\rm tot}})$ provides a QND detection of $(i\gamma_1\gamma_2)(i\gamma_3\gamma_4)$.

\emph{\blue{Teleportation}} -- The overscreening in the odd-$N$ case can be used to coherently transfer entanglement from one Kondo anyon to the other. Starting with a $J_R=0$ system, initialized such that the $\gamma_L$ MZM is entangled with an ancillary MZM $\eta$ and the detached spin is polarized in the $S^z_R=-1/2$ direction, we adiabatically turn on $J_R$ and turn off $J_L$, as shown in Fig.\,\ref{Fig3}(a) for $N=1$. At the end of the process $\eta$ is entangled with $\gamma_R$, equivalent to a transfer of the MZM from one spin to the other spin, while coherently maintaining its entanglement with the ancillary MZM. This operation is guaranteed by the fermion parity conservation ($\sigma_L\to\sigma_R$ and $\tau_L\to\tau_R$), while the backflow of spin $\vec S_R\to\vec S_L$ is protected by the conserved $\vec{\cal I}$ charge. This is reminiscent of (but different from) the electron teleportation effect \cite{Semenoff2006,Bolech2007,Fu2010,Vijay2016,Whiticar2020,Huang2021c,Zhang2023e}. In longer odd-$N$ chains, it is sufficient to change the tunnel couplings while maintiaining $J_L=J_R$ \cite{SM}.

\emph{\blue{Switching across a spin}} -- Likewise, the entangled MZM can be moved across a local moment  [Fig.\,\ref{Fig3}(b,c)]. However the two sides have independent fermion parities, and therefore, a small tunnel coupling $t_{xy}$ is needed to break the independent gauge invariance of the two sites. It can be shown that, up to normalization factors \cite{SM},
\vspace{-.1cm}
\be
\mat{\gamma_L \\ \gamma_R}\to\mat{\cos2\vartheta & \sin2\vartheta \\ -\sin2\vartheta&\cos2\vartheta}\mat{\gamma_L \\ \gamma_R},\vspace{-.1cm}
\ee
with the angle $\vartheta=\int_{t_i}^{t_f}t_{xy}(t)[\cos2\alpha(t)-1/2]$, depending on the details of pulse sequence, where $\tan2\alpha=\sqrt3(J_L-J_R)/(J_L-J_R)$. This is clearly a non-topological operation (unlike the teleportation) \cite{SM}, but it is easy to adjust $t_{xy}$ to get $\vartheta=\pm \pi/4$ for positive/negative $t_{xy}$ amplitudes, corresponding to a direct [\ref{Fig3}(c)] or sign-flipped [\ref{Fig3}(b)] transfer of entanglement. In this case, the operation can be represented as $R_{LR}$ or $R_{RL}$ where
\vspace{-.1cm}
\be
R_{ab}\equiv \frac{1}{\sqrt 2}(1+2\gamma_a\gamma_b).\vspace{-.1cm}
\ee

\begin{figure}[tp]
\centering
\includegraphics[width=\linewidth]{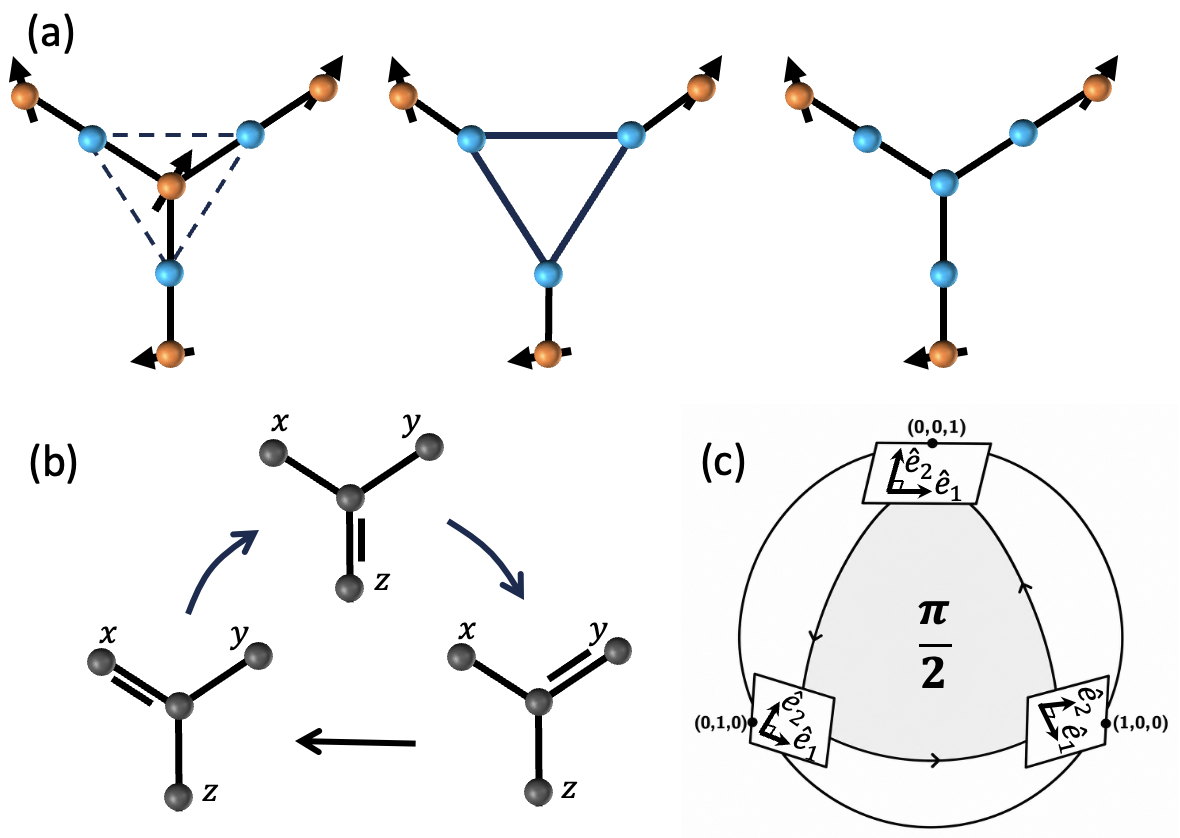}
\caption{\small\raggedright (a) Setups considered for braiding; $S$-centered, $c$-triangle, and $c$-centered Y-junction. (b) Braiding protocol applied to non-interacting particles on a c-centered Y-junction. (c) The parallel transport of dark fermions over a unit sphere enclosing a $4\pi/8$ solid angle results in a robust holonomy producing topological braiding.\vspace{-.4cm}}
\label{Fig4}
\end{figure}

\emph{\blue{Braiding}} requires a minimal two-dimensional geometry, naturally realized by the Y-junctions shown in Fig.\,\ref{Fig4}(a) \cite{Ftnote1}. In all cases, we initialize two detached Kondo MZMs on the $x$ and $y$ legs and perform the cyclic transfer sequence
$y\to z$, $x\to y$, and $z\to x$, thereby exchanging the two Kondo anyons [Fig.\,\ref{Fig4}(b)].

The nature of the resulting operation depends crucially on the junction geometry. In the $S$-centered junction, transferring a Kondo MZM across the central spin requires additional tunneling terms. The resulting transport is non-topological, depends on pulse design, and cannot be generated from gauge-invariant couplings in the original model. Although the sequence reproduces the exchange operator $R_{xy}$, the operation itself is not protected and therefore does not constitute a genuine topological braid.

The triangle-centered junction likewise fails to realize a topological exchange: for even fermion parity on each leg teleportation is absent, while for odd parity the transfer remains classical and does not support braiding.

By contrast, the $c$-centered Y-junction realizes a genuinely topological braid operation. In this geometry the exchange emerges entirely from adiabatic transport through a degenerate dark-state manifold, producing a robust non-abelian Berry curvature discussed below.

\emph{\blue{Wilczek-Zee holonomy}} -- Consider the Hamiltonian $H(t)=c\dg_0(\vec v\cdot\vec c)+h.c.$, with $\vec v=(v_x,v_y,v_z) $. This non-interacting Y-junction provides a geometric scaffold that survives dressing by Kondo interactions.
At any given time $c_0$ only couples to a bright mode $b=\hat v\cdot\vec c$ with $\hat v=\vec v/\vert\vec v\vert$, forming a gapped pair, while two dark (zero energy) modes 
 $d_1=\hat e_1\cdot\vec c$ and $d_2=\hat e_2\cdot\vec c$ are decoupled.  The orthonormal triad ${\cal U}(t)=(\hat v,\hat e_1,\hat e_2)$ diagonalizes the instantaneous Hamiltonian and defines a two-dimensional dark-state manifold tangent to the unit sphere traced by $\hat v$. Adiabatic evolution within this degenerate manifold is governed by the projected connection $A=P_d({\cal U}\dg d\,{\cal U})P_d$, which takes the form $A=i\sigma^y\omega$ with $d\omega$ the area 2-form on the unit sphere. Restricting the evolution such that one tunneling amplitude vanishes at each stage, the closed trajectory encloses a quantized solid angle and produces a geometric holonomy \cite{SM}
\vspace{-.1cm}
\be
U\equiv{\cal P}\exp\big[-\oint \vec A\cdot d\vec\ell\,\,\big]=\exp[-{\frac{i\pi}{2}\sigma^y}]=-i\sigma^y.
\vspace{-.1cm}
\ee
The resulting evolution $\hat U\dg \vec c\,\hat U=U\vec c$ yields the holonomic braid transformation $c_x\to -c_y$ and $c_y\to c_x$. 

The braid operation therefore emerges as a Wilczek--Zee holonomy \cite{Wilczek1984} associated with parallel transport through a dark-state manifold, connecting Majorana braiding \cite{Alicea2011} to geometric pumping and adiabatic-passage phenomena \cite{Thouless1983,Brouwer1998,Vitanov2017}. In this formulation, the braid structure is fundamentally geometric and does not rely on microscopic branch-cut constructions \cite{Ivanov01} or even the statistics of the underlying particles \cite{Ftnote3}. What distinguishes these Majorana modes is the robustness of the encoded non-local information against local perturbations.

\emph{\blue{Kondo dressing}} --  The same geometric structure survives dressing by Kondo interactions, leading to
$\vec S_x\lr \vec S_y$ and $\vec S_z\to \vec S_z$. The resulting operation is $R_{xy}$, producing the non-Abelian exchange $\gamma_x\to-\gamma_y$ and $\gamma_y\to\gamma_x$ of Kondo MZMs \cite{SM}.
Unlike free-fermion MZMs, however, $\gamma$ modes remain accompanied by nearby $\chi^0$ modes, and their decoupling is protected only by channel symmetry. 

\emph{\blue{Scalability and qubit operations}} -- 
Kondo MZMs can be used in both sparse and dense qubit encodings \cite{Sarma2015,Beenakker2020,Frey2026}. Isolated $\gamma$ modes are accompanied by their own $\chi^0$ modes, naturally realizing sparse encoding. In multi-impurity setups, however, several $\gamma$ modes can share the same $\chi^0$, enabling dense encoding. The two Majoranas $\gamma_x$ and $\gamma_y$ of a Kondo qubit can be coupled through a shared $\chi^0$ by tuning channel asymmetry between $J_x$ and $J_y$, thereby enabling arbitrary single-qubit operations. Furthermore, the $c$-centered Y-junction naturally generalizes to networks with multiple Kondo legs, such as radial or Cayley-tree geometries.

Braiding can be utilized to realize not only a topological Hadamard gate, but also two-qubit operations like the CNOT gate \cite{Barenco1995,Boykin1999,Georgiev2006,Bravyi2006,vanHeck2012,Nielsen2012,Beenakker2020,Zhang2024d,SM}. Alternatively, two-qubit operations can be achieved through QND parity measurement \cite{Beenakker2004,Zilberberg2008,Beenakker2020,SM}.

\emph{\blue{Experimental considerations}} -- 
While experimentally realized SET-IQH charge-Kondo platforms \cite{Iftikhar2015,Iftikhar2018,Pouse2023} provide a natural motivation for the present framework, they differ microscopically from the multi-impurity setups studied here. In particular, only part of the edge degrees of freedom are shared between impurities, and establishing a controlled mapping to the effective models considered in this work requires further case-specific microscopic analysis.

A second challenge is the presence of gapless bulk excitations, which impose stringent constraints on adiabatic operations. One possible route, proposed in Ref.\,\cite{Komijaniqubit}, is to gap the bulk through superconducting proximity while preserving the Kondo Majorana zero modes \cite{Komijaniqubit,Hur2000}. Realizing such architectures experimentally remains an important open problem.

\emph{\blue{Summary}} -- 
Our results establish a proof-of-principle framework for interacting non-Abelian anyons beyond free-fermion platforms. We show that 2CK impurities support emergent Majorana zero modes that can store information in non-local qubits and undergo coherent teleportation, fusion, and braiding operations. In particular, we identify a distinction between non-topological and genuinely topological Y-junction geometries, the latter realizing a non-Abelian geometric holonomy. Beyond their implications for Ising anyons, these results suggest that multi-channel Kondo platforms may provide an 
experimentally accessible route toward more exotic anyonic structures beyond non-interacting architectures.

\emph{\blue{Acknowledgements}} --  CJB gratefully acknowledges financial support for part of this work by the Fulbright U.S. Scholar Program, funded by the US-DoS and Fulbright España; (the contents of this work do not represent the official views of these institutions).

\bibliography{Kondoanyons3}

\begin{thebibliography}{62}
\bibitem [{\citenamefont {{We use $\dul\sigma^a$ and (later) $\dul\tau^a$
  symbols to refer to Pauli matrices in pseudospin and isospin spaces,
  respectively, while $\ket{\sigma^a}$ and $\ket{\tau^a}$ symbols will be used
  to refer to the pseudospin and isospin singlet and triplet
  states}}()}]{Ftnote2}%
  \BibitemOpen
  \bibfield  {author} {\bibinfo {author} {\bibnamefont {{We use $\dul\sigma^a$
  and (later) $\dul\tau^a$ symbols to refer to Pauli matrices in pseudospin and
  isospin spaces, respectively, while $\ket{\sigma^a}$ and $\ket{\tau^a}$
  symbols will be used to refer to the pseudospin and isospin singlet and
  triplet states}}},\ }\href@noop {} {}\BibitemShut {NoStop}%
\end{thebibliography}

\newpage
\newpage

\bw

\section*{\large Supplementary Materials}

\setcounter{secnumdepth}{2} 
\setcounter{tocdepth}{2}    

\makeatletter

\renewcommand{\l@section}[2]{%
  \addpenalty{\@secpenalty}%
  \addvspace{1.0em plus\p@}%
  \begingroup
    \parindent \z@
    \rightskip \@pnumwidth
    \parfillskip -\@pnumwidth
    \leavevmode
    \bfseries
    #1\nobreak
    \leaders\hbox{$\m@th
      \mkern \@dotsep mu\hbox{.}\mkern \@dotsep mu$}%
    \hfill
    \nobreak\hb@xt@\@pnumwidth{\hss #2}\par
  \endgroup
}

\renewcommand{\l@subsection}[2]{%
  \begingroup
    \parindent 1.5em
    \rightskip \@pnumwidth
    \parfillskip -\@pnumwidth
    \leavevmode
    #1\nobreak
    \leaders\hbox{$\m@th
      \mkern \@dotsep mu\hbox{.}\mkern \@dotsep mu$}%
    \hfill
    \nobreak\hb@xt@\@pnumwidth{\hss #2}\par
  \endgroup
}

\renewcommand{\l@subsubsection}[2]{}

\setcounter{figure}{0}
\setcounter{equation}{0}
\renewcommand{\thefigure}{S\arabic{figure}}
\renewcommand{\theequation}{S\arabic{equation}}

\makeatother

\title{Supplementary Materials for \\
Coherent manipulations of Kondo Majoranas in two-channel Kondo setups}
\author{Yashar Komijani\,$^{*}$}
\author{Carlos Bolech}
 \affiliation{ Department of Physics, University of Cincinnati, Cincinnati, Ohio, 45221, USA}
\date{\today}
\maketitle
This supplementary material provides proof of various statements made in the paper including an explicit computation of the beta function coefficients.
\tableofcontents
\section{\large\sc Mapping to the compactified chain}
\subsection{Mapping to the compactified continuum model}
In this section we use bosonization/fermionization technique to derive a minimal version of the model, involving only two 
chiral fermions. The problem of a single Kondo impurity spin coupled to two spinful channels (edge states) contains 4 complex fermions, $\psi_{a\sigma}$ where $a=1,2$ are channels and $\sigma=\ua,\da$ are the spins.

For concreteness, we consider a chiral (left-mover) circular edge state $x\in(-L,L)$ (with periodic or anti-periodic boundary condition) which is spinful and has two channels,
\be
H_0=iv_F\sum_{a=1,2}\sum_{\alpha=\ua,\da}\int_{-L}^Ldx \psi\dg_{a\alpha}\partial_x\psi_{a\alpha}.
\ee
We consider a model in which multiple impurities are coupled to the same chiral edge states \cite{Ftnote2}
\be
H=H_0+\sum_{j=1}^N\vec S_j\cdot 
\sum_{a=1}^2J_K^{(a)}\psi\dg_{a\alpha}(x_j)[\vec{\dul\sigma}]_{\alpha\beta}\psi\dn_{a\beta}(x_j).
\ee
Through bosonization, and re-fermionization, we map the multi-impurity 2CK model \cite{Ljepoja2024A,Ljepoja2024B,Ljepoja2024C}.  Unlike Toulouse-point constructions based on the Emery–Kivelson transformation \cite{Emery1993}, the compactified representation does not rely on fine-tuning to a solvable point, but instead provide an economical strong-coupling description of the generic 2CK fixed point. Therefore, we work with spin-isotropic Kondo couplings throughout in an extended form
\be
H=H_0+\sum_{j}\sum_{a=1,2} J^{(a)}_j\Big\{[S_j^-\psi\dg_{a\ua}(x_j)\psi\dn_{a\da}(x_j)+h.c.]+S_j^z[\psi\dg_{a\ua}(x_j)\psi\dn_{a\ua}(x_j)-\psi\dg_{a\da}(x_j)\psi\dn_{a\da}(x_j)]\Big\}.
\ee

The key simplification is that the impurity couples only to spin and spin-flavor sectors, while charge and flavor sectors decouple completely. We bosonize $\psi_{\sigma}=\frac{1}{\sqrt{2\pi a}}F_\sigma e^{i\phi_{\sigma}}$ with the short distance UV cut off $a$. These bosons satisfy [$\alpha,\beta=\ua,\da$ spins and for $a,b=1,2$ channels]:
\be
[\phi_{a\alpha}(x),\partial_y\phi_{b\beta}(y)]=-2\pi i \delta(x-y)\delta_{ab}\delta_{\alpha\beta}.
\ee
We do an orthogonal transformation of the original bosons to a set of collective mode bosons
\bea
\mat{\phi_c \\ \phi_s \\ \phi_f \\ \phi_{s\! f}}=\frac{1}{2}\matc{cccc}{
1 & 1 & 1 & 1 \\
1 & -1 & 1 & -1 \\
1 & 1 & -1 & -1 \\
1 & -1 & -1 & 1
}
\mat{\phi_{1\ua} \\ \phi_{1\da} \\ \phi_{2\ua} \\ \phi_{2\da}}.
\eea
Using the identities
\be
F\dg_{1\ua}F\dn_{1\da}=F\dg_sF\dg_{s\! f}, \qquad 
F\dg_{2\ua}F\dn_{2\da}=F\dg_sF\dn_{s\! f},
\ee
we can write
\bea
H&=&H_0+\frac{1}{{2\pi a}}\sum_j \Big\{\Big(S^-_jF\dg_se^{-i\phi_s(x_j)}[J^{(1)}_j F\dg_{s\! f}e^{-i\phi_{s\! f}(x_j)}+J^{(2)}_jF\dn_{s\! f}e^{+i\phi_{s\! f}(x_j)}]+h.c.\Big)\\
&&\hspace{6cm}+\frac{1}{\sqrt{2\pi}}S^z_j\Big[(J_j^{(1)}+J_j^{(2)})\partial_x\phi_s(x_j)+(J^{(1)}_j-J^{(2)}_j)\partial_x\phi_{s\! f}(x_j)\Big]\Big\}.
\eea
We now refermionize the spin and spin-flavor sectors through $\psi_\mu=\frac{1}{\sqrt{2\pi a}}F_\mu e^{i\phi_\mu}$.
\bea
H&=&H_0+\sum_j \Big\{\Big(S^-_j\psi\dg_s(x_j)[J^{(1)}_j \psi\dg_{s\! f}(x_j)+J^{(2)}_j\psi\dn_{s\! f}(x_j)]+h.c.\Big)\\
&&\hspace{6cm}+S^z_j\Big[(J_j^{(1)}+J_j^{(2)})\psi\dg_s(x_j)\psi\dn_s(x_j)+(J_j^{(1)}-J_j^{(2)})\psi\dg_{s\! f}(x_j)\psi\dn_{s\! f}(x_j)\Big]\Big\}.\quad
\eea
where
\be
H_0=iv_F\int_{-L}^Ldx[\psi\dg_c\partial_x\psi\dn_c+\psi\dg_f\partial_x\psi\dn_f+\psi\dg_s\partial_x\psi\dn_s+\psi\dg_{s\! f}\partial_x\psi\dn_{s\! f}].
\ee
Renaming $\psi_s\to\tilde\psi_\uparrow$ and $\psi_{s\! f}\to\tilde\psi_\downarrow$, as two spin species, satisfying the boundary conditions
\be
\tpsi_{\ua}(L)=\tpsi_{\ua}(-L), \qquad \tpsi_{\da}(L)=\tpsi_{\da}(-L).
\ee
Dropping the decoupled charge $\psi_c$ and flavor $\psi_f$, the free part of the Hamiltonian becomes
\be
H_0\to iv_F\int_{-L}^L[\tpsi\dg_\ua\partial_x\tpsi_\ua+\tpsi\dg_{\da}\partial_x\tpsi_{\da}]dx
\ee
and the rest of the Hamiltonian is
\bea
H=H_0+\sum_j\Big\{&&J_j^{(2)}\Big(S^-_j[\tpsi\dg_\ua(x_j)\tpsi\dn_\da(x_j)+h.c.]+S_j^z[\tpsi\dg_\ua(x_j)\tpsi\dn_\ua(x_j)-\tpsi\dg_\da(x_j)\tpsi\dn_\da(x_j)]\Big)\\
&&\hspace{3cm}+J^{(1)}_j\Big(S^-_j[\tpsi\dg_\ua(x_j)\tpsi\dg_\da(x_j)+h.c.]+S_j^z[\tpsi\dg_\ua(x_j)\tpsi\dg_\ua(x_j)+\tpsi\dg_\da(x_j)\tpsi\dn_\da(x_j)]\Big)\Big\}.\quad
\eea
Collecting $\tilde\psi_\ua$ and $\tilde\psi_\da$ in the Nambu spinor $\tilde\Psi$, the problem is reduced to the so-called compactified model \cite{Coleman1995,Bulla1997b,Bulla1997} in which the interaction is of the form  
\be
H=H_0+\frac{1}{2}\sum_j\vec S_j\cdot \tilde\Psi\dg(x_j)[J^{(1)}_j\vec{\dul\tau}+J_j^{(2)}\vec{\dul\sigma}]\tilde\Psi(x_j)\quad \tilde\Psi\dg=\matc{cccc}{\tpsi\dg_\ua & \tpsi\dg_\da & \tpsi\dn_\da & -\tpsi\dn_\ua}. \label{eq15}
\ee
In this compactified form, the  $\vec\sigma$ and $\vec\tau$ act in spin and (particle-hole) isospin degrees of freedom of a single-channel of fermions. 
\subsection{Mapping to the tight binding chain}
For concreteness we focus on the two-impurity two-channel Kondo problem. We show that the tight-binding Hamiltonian \cite{compactified} which contains a two-impurity compactified two-channel Kondo model
\bea
H&=&-t\sum_{n=1}^N \sum_{\alpha=\ua,da}(c\dg_{n,\alpha}c\dn_{n+1,\alpha}+h.c.)+\frac{1}{2a}\Big[\vec S_1\cdot C_1\dg(J_L^{(1)}\vec{\dul\tau}+J_L^{(2)}\vec{\dul\sigma})C_1+\vec S_2\cdot C\dg_N(J_R^{(1)}\vec{\dul\tau}+J_R^{(2)}\vec{\dul\sigma})C_N\Big],
\eea
with
\be
C\dg\equiv\matc{cccc}{c\dg_\ua & c\dg_\da & c\dn_\da & -c\dn_\ua},
\ee
has the same long-distance behavior as Eq.\,\pref{eq15}. The first term is a tight-binding chain of a single-channel spinful electron at half-filling with $k_F=\pi/2$. Furthermore, $\vec{\dul\sigma}$ and $\vec{\dul\tau}$ are two sets of commuting Pauli matrices $[\dul\sigma^a,\dul\tau^b]=0$, obeying the commutation algebras $[\dul\sigma^a\,\dul\sigma^b]=2i\eps^{abc}\dul\sigma^c$, and $[\dul\tau^a,\dul\tau^b]=2i\eps^{abc}\dul\tau^c$. 

In the continuum limit, we can expand the fermions in terms of right- and left-movers around the Fermi energy:
\be
\frac{1}{\sqrt a}c\dn_{n,\alpha}=i^n\psi_{R,\alpha}(na)-(-i)^n\psi_{L,\alpha}(na).
\ee
These operators (and the corresponding wavefunctions) $c_{j,\sigma}$ vanish at the phantom sites $j=0$ and $j=N+1$.  This leads to boundary conditions
\be
\psi_{R,\alpha}(0)=\psi_{L,\alpha}(0), \quad \psi_{R,\alpha}(L)=(-1)^{N+1}\psi_{L,\alpha}(L).
\ee
where $L=Na$. We can work with odd-$N$ chains, with periodic boundary condition at $x=L$ for the unfolded chiral fermion, or even $N$ chains with anti-periodic boundary condition at $x=L$. The original circular edge state problem can access both cases by encircling an appropriate Aharonov-Bohm flux. To be concrete, we choose periodic boundary conditions and unfold the system to only left-mover
\be
\tilde\psi_\alpha(x)\equiv\left\{\matl{\psi_{L,\sigma}(x) & \quad x\in (0,L) \\ \psi_{R,\sigma}(-x) &\quad x\in (-L,0)}\right.
\ee
obeying $\tilde\psi_\alpha(-L)=\tilde\psi_\alpha(L)$. Then naturally
\be
\frac{1}{\sqrt a}C_1=\tilde\Psi(0), \andd \frac{1}{\sqrt a} C_N=\tilde\Psi(L),
\ee
so that the Hamiltonian becomes
\bea
H=iv_F\int_{-L}^Ldx[\tpsi\dg_\ua \partial_x\tpsi_\ua+\tpsi_\da\dg\partial_x\tpsi\dn_\da]
+\frac{1}{2}\Big\{\Big[\vec S_1\cdot \tilde\Psi\dg(0)[J_L^{(1)}\vec{\dul\tau}+J_L^{(2)}\vec{\dul\sigma}]\tilde\Psi(0)\Big]+
\Big[\vec S_2\cdot \tilde\Psi\dg(L)[J^{(1)}_R\vec{\dul\tau}+J_R^{(2)}\vec{\dul\sigma}]\tilde\Psi(L)\Big]\Big\},\qquad
\eea
which is precisely Eq.\,\pref{eq15} with equidistant impurities.
\section{\large\sc Exact strong-coupling analysis}
In this section we show that\\
(i) the strong-coupling manifold carries a localized composite Majorana;\\
(ii) adiabatic S-c-S, S-c-c-c-S and c-S-c protocols transport the projected Majorana zero mode;\\
(iii) the c-S-c transport acquires a controllable even--odd dynamical phase that depends on the tunneling amplitude between the two $c$-s. \\
(iv) The c-centered Y junction can be used for braiding both non-interacting particles and Kondo MZMs.\\
(v) The S-c-c-S setup can be used to fuse Kondo MZMs.
\subsection{Symmetries in S-c setups}
We consider the usual compactified two-channel Kondo (c2CK) model with a single impurity-spin
\be
H=H_0+\frac{1}{2}\vec S\cdot C\dg(J\vec{\dul\sigma}+J'\vec{\dul\tau})C, \qquad C=(\bmx{cccc} c\dn_\ua & c\dn_\da & c\dg_\da & -c\dg_\ua \emx)^T.
\ee
The model can be written as
\be
H=H_0+\vec S\cdot (J^{(2)} c\dg\vec{\dul\sigma} c+J^{(1)}\tilde c\dg\vec{\dul\sigma} \tilde c), \qquad c=(\matn{c_\ua & c_\da})^T \qquad \tilde c=(\matn{c_\ua & c\dg_\da})^T 
\ee
As before, $[\dul\sigma^a\,\dul\sigma^b]=2i\eps^{abc}\dul\sigma^c$ and $[\dul\tau^a,\dul\tau^b]=2i\eps^{abc}\dul\tau^c$ with $[\dul\sigma^a,\dul\tau^b]=0$. We have
\be
c\dg\vec\sigma c=\Big\{(c\dg_\ua c\dn_\da+h.c.),-i(c\dg_\ua c\dn_\da-h.c.),(c\dg_\ua c\dn_\ua-c\dg_\da c\dn_\da)\Big\},\quad
\tilde c\dg\vec\sigma \tilde c=\Big\{(c\dg_\ua c\dg_\da+h.c.),-i(c\dg_\ua c\dg_\da-h.c.),(c\dg_\ua c\dn_\ua+c\dg_\da c\dn_\da-1)\Big\}.\qquad
\ee
With $\sqrt{2}c\dg_\sigma=\chi'_\sigma+i\chi''_\sigma$ we have 
\be
\vec{\cal J}_0\equiv\frac{1}{2}(c\dg\vec{\dul\sigma} c+\tilde c\dg\vec{\dul\sigma} \tilde c)=i\{\chi''_\ua \chi'_\da, \chi'_\da \chi'_\ua,\chi''_\ua \chi'_\ua\},\qquad \vec{\cal J}_0'\equiv\frac{1}{2}(c\dg\vec{\dul\sigma} c-\tilde c\dg\vec{\dul\sigma} \tilde c)=i\{\chi'_\da\chi_\ua,\chi'_\ua\chi'_\da,\chi_\da\chi'_\da\},
\ee
in terms of which
\be
H=H_0+\vec S\cdot (\bar J\vec {\cal J}+\delta J\vec{\cal J}'),
\ee
where $\bar J=(J^{(2)}+J^{(1)})/2$ and $\delta J=J^{(2)}-J^{(1)}$. We define  $(\chi^x,\chi^y,\chi^z,\chi^0)=(-\chi'_\ua,-\chi''_\ua, \chi'_\da,\chi''_\da)$ so that
\be
\vec{\cal J}=-i\Big\{\chi^y\chi^z,\chi^z\chi^x,\chi^x\chi^y\Big\}, \andd 
\vec{\cal J}'=-i\chi^0\Big\{\chi^x,\chi^y,\chi^z\Big\}. 
\ee
Note that while the elements of $\vec{\cal J}$ obey an SU(2) algebra with ${\cal J}_0^2=3/4$, $\vec{\cal J}'$ does not close under SU(2):
\be
[{\cal J}^a,{\cal J}^b]=\frac{i}{2}\eps^{abc}{\cal J}^c,\qquad [{\cal J}{'}^a,{\cal J}{'}^b]=\frac{i}{2}\eps^{abc}{\cal J}^c,\andd [{\cal J}^a,{\cal J}{'}^b]=
\frac{i}{2}\eps^{abc}{\cal J}{'}^c.
\ee
These two currents do not commute with each other. With the proper choice
\be
H_0=\sum_j(itc\dg_jc\dn_{j+1}+h.c.)=it\sum_j\chi_j^\mu\chi_{j+1}^\mu, \qquad \mu=x,y,z,0.
\ee
Defining the currents
\be
{\cal J}^a_n=-\frac{i}{2}\eps^{abc}\chi_n^b\chi_n^c,\qquad \vec{\cal J}_{tot}=\sum_n\vec{\cal J}_n,
\ee
we have
\be
\vec{\cal J}_{tot}^2=\frac{3N}{4}+2\sum_{n<m}\vec{\cal J}_n\cdot\vec{\cal J}_m.
\ee
It can be shown that
\be
[\vec{\cal J}_{tot},H_0]=0, 
\ee
and therefore, 
\be
[\vec S+{\cal J}_{tot},H]=0,
\ee
which also implies $[(\vec S+{\cal J}_{tot})^2,H]=0$.
\subsection{Kondo MZM in S-c setup}
Although $\vec{\cal J}$ needs three MZMs, such a Hilbert space has a $Z_2$ anomaly as the fermion parity cannot be defined. However, with four MZMs, it has two representations:
\be
\text{even:}\quad \Big\{\ket{0},\ket{\ua\da}\Big\} \so \frac{1}{2}\Big\{\dul\sigma^x,\dul\sigma^y,\dul\sigma^z\Big\}, \andd
\text{odd:}\quad \Big\{\ket{\da},\ket{\ua}\Big\} \so \frac{1}{2}\Big\{\dul\sigma^x,\dul\sigma^y,\dul\sigma^z\Big\}. \qquad
\ee
For a c-S setup the Hamiltonian is $H=\vec S\cdot\vec{\cal J}$. This Hamiltonian can be written as
\be
H=\frac{1}{2}[(\vec S+\vec{\cal J})^2-\vec S^2-\vec{\cal J}^2]=\left\{\matl{1/4 & j=1 \\ -3/4 & j=0}\right.
.
\ee
The even/odd parity sectors (adding back the decoupled $\chi''_\da$) are
\be
{\cal H}^{odd}=\Big\{\ket{\sigma}=\frac{\ket{\Ua\da}-\ket{\Da\ua}}{\sqrt 2},\quad \ket{\vec \sigma}=\big(\ket{\Ua\ua},\frac{\ket{\Ua\da}+\ket{\Da\ua}}{\sqrt 2},\ket{\Da\da}\big)\Big\},\label{eqS35}
\ee
and
\be
{\cal H}^{even}=\Big\{\ket{\tau}=\frac{\ket{\Ua0}-\ket{\Da2}}{\sqrt 2},\quad \ket{\vec \tau}=\big(\ket{\Ua2},\frac{\ket{\Ua0}+\ket{\Da2}}{\sqrt 2},\ket{\Da0}\big)\Big\}.\label{eqS36}
\ee
Now, defining $\gamma=Z\vec S\cdot\vec \chi$, we find that $\gamma^2=Z^2(\frac{3}{8}-H)$ is directly related to $H$. Demanding $\gamma^2\sim \bb 1/2$ on the ground state (GS) manifold with energy $H=-3/4$, we choose $Z=2/3$. This means that on the singlet sector $\gamma^2=1/2$ but on the excited state triplet sector $\gamma^2=1/18$. It can be explicitly checked that, e.g. ,
\be
{\sqrt 2}\gamma:\qquad \ket{\sigma}\to\ket{\tau}, \qquad \ket{\tau}\to\ket{\sigma},\qquad 
\ket{\vec\sigma}\to\frac{1}{3}\ket{\tau}, \qquad \ket{\vec\tau}\to\frac{1}{3}\ket{\vec\sigma}. 
\ee
Although the projected $\gamma$ operator is not an exact microscopic Majorana fermion, it closes faithfully within the low-energy manifold. A similar property is obeyed by the decoupled $\chi^0$ (not only on the GS manifold), meaning that ${\chi^0}^2=1/2$ and
\be
\sqrt{2}\chi^0:\qquad \ket{\sigma}\to i\ket{\tau}, 
\quad
\ket{\tau}\to -i\ket{\tau},
\qquad \ket{\vec\sigma}\to i\ket{\vec\tau},\quad \ket{\vec\tau}\to -i\ket{\vec\sigma},
\ee
which, unlike $\gamma$, has trivially kept its normalization intact even when acting on the excited states.

\subsection{Coupling to an auxiliary MZM}
Suppose we have the Hamiltonian (with $0<m\ll J_K$)
\be
H=im\gamma\eta',
\ee
defined by adding a fermionic state $\sqrt{2}f\dg=\eta'+i\eta''$. In total, there are 4 low energy states with opposite parities
\be
\ket{\pm}^e=\frac{\ket{\tau}\ket{0}_f\mp i\ket{\sigma}\ket{1}_f}{\sqrt{2}}, \andd \ket{\pm}^o=\frac{\ket{\sigma}\ket{0}_f\pm i\ket{\tau}\ket{1}_f}{\sqrt{2}}.
\ee
Their energies are $E_\pm^{e/o}=\mp m/2$.

\subsection{The S-c-S setup}
Next we consider the S-c-S geometry, with the Hamiltonian
\be
H=\vec{\cal J}\cdot(J_L\vec S_L+J_R\vec S_R)=\vec{\cal J}\cdot\big(\bar J\vec L+\frac{1}{2}{\delta J}\vec M \big),
\ee
where $\bar J=(J_L+J_R)/2$ and $\delta J=J_L-J_R$, and we have defined
\be
 \vec L=\vec S_L+\vec S_R, \qqquad \vec M=\vec S_L-\vec S_R.
\ee
For $\delta J=0$ the eigenstates are
\be
H/\bar J=\vec{\cal J}\cdot\vec L=\frac{1}{2}[j(j+1)-\ell(\ell+1)-3/4]=\left\{\matl{1/2 &\ell=1, j=3/2 \\ 0 & \ell=0,j=1/2 \\  -1 & \ell=1, j=1/2}\right.
.
\ee
Note that ${\cal J}$ has two parity sectors (re-including $\chi^0=\chi''_\da$). We denote eigenstates $\ket{\ell,m_\ell}$ between two spins with singlet $\ket{S}$ and triplet $\ket{T_+}$, $\ket{T_-}$ and $\ket{T_0}$ symbols. In terms of these, the over-screened eigenstates $\kett{j,m_j}$, with the right Clebsch-Gordan coefficients in the odd/even sectors are
\bea
\kett{1/2,1/2}^{o/e}\equiv\sqrt{2/3}\ket{T_+}\ket{\da/0}-\sqrt{1/3}\ket{T_0}\ket{\ua/2}, \quad \kett{1/2,-1/2}^{o/e}\equiv\sqrt{1/3}\ket{T_0}\ket{\da/0}-\sqrt{2/3}\ket{T_-}\ket{\ua/2}.\quad
\eea
Here, the $o$ superscript on the left involves $\ket{\ua/\da}$ states on the right, whereas the $e$ superscript involves $\ket{2/0}$ states on the right. For future references, we also note the low-lying excited states in $\ell=0$ sector, with energy $H=0$,
\be
\kett{\ell=0,1/2}^o=\ket{S}\ket{\ua}, \quad \kett{\ell=0,-1/2}^o=\ket{S}\ket{\da}, \quad 
\kett{\ell=0,1/2}^e=\ket{S}\ket{2}, \quad \kett{\ell=0,-1/2}^e=\ket{S}\ket{0}.
\ee
The $j=3/2$ state is not reached from the GS. When $\delta J\neq 0$
\be
\hat V/(\delta J/2)=M^z{\cal J}^z+\frac{1}{2}(M^+{\cal J}^-+M^-{\cal J}^+), \qquad {\cal J}^+=c\dg_\ua(c\dg_\da+c\dn_\da), \quad 
{\cal J}^-=(c\dg_\da+c\dn_\da)c\dn_\ua
\quad {\cal J}^z=c\dg_\ua c\dn_\ua-1/2.
\ee
Considering that
\be
\hspace{-.1cm}-M^-\ket{T_+}=M^+\ket{T_-}={\sqrt 2}M^z\ket{T_0}=\sqrt{2}\ket{S}, \quad 
M^z\ket{S}=\ket{T_0}, \quad M^+\ket{S}=-\sqrt{2}\ket{T_+},\quad M^-\ket{S}=\sqrt{2}\ket{T_-}.
\ee
we find  $V=-\sqrt{3}\delta J/4$ so that
\be
\hat V\kett{1/2,1/2}^o=V\ket{S}\ket{\ua}, \quad \hat V\kett{1/2,-1/2}^o=V\ket{S}\ket{\da}, \quad \hat V\kett{1/2,1/2}^e=V\ket{S}\ket{2},\quad
\hat V\kett{1/2,-1/2}^e=V\ket{S}\ket{0}.
\ee
This results in a 2x2 system with level-repulsion, and the eigenvalues
\be
H=\bar J\mat{-1 \\ & 0}+\mat{& V \\ V &}, \qquad  E_\pm=-\frac{1}{2}\bar J\pm \sqrt{(\bar J/2)^2+V^2}=-\frac{1}{2}\bar J\pm \sqrt{(\bar J/2)^2+\frac{3}{16}\delta J^2}.
\ee
Thus, we find that the GS with energy $E_-$ is given by [using $\tan2\alpha=2{V}/{\bar J}$ and  $V=-\sqrt{3}\delta J/4$]
\bea
\kettt{\Ua}^{o/e}\equiv\cos\alpha\kett{1/2,1/2}^{o/e}-\sin\alpha\ket{S}\ket{\ua/2},\qquad
\kettt{\Da}^{o/e}\equiv\cos\alpha\kett{1/2,-1/2}^{o/e}-\sin\alpha\ket{S}\ket{\da/0}.
\eea
Again, the odd sector gets admixture from the states $\ket{\ua/\da}$ and the even sector gets admixture from the states $\ket{2/0}$. So, as we go from $J_R=0$ ($\alpha=-\pi/6$) to $J_L=0$ ($\alpha=\pi/6$), the angle $\alpha(t)$ varies in time. For future references we also have
\bea
\kettt{\Ua'}^{o/e}\equiv\sin\alpha\kett{1/2,1/2}^{o/e}+\cos\alpha\ket{S}\ket{\ua/2},\qquad
\kettt{\Da'}^{o/e}\equiv\sin\alpha\kett{1/2,-1/2}^{o/e}+\cos\alpha\ket{S}\ket{\da/0},
\eea
which are the excited states, with energies $E_+$.
\subsubsection{Matrix elements of $\gamma_L$ and $\gamma_R$}
In this subsection we show that the projected Majorana operators close within the low-energy manifold and interpolate continuously between left- and right-localized Kondo Majoranas. When $J_R=0$ the GS subspace is
\bea
\kettt{\Ua_R}^o&=&\ket{\sigma_L}\ket{\Ua_R}=\frac{\sqrt 3}2\kett{1/2,1/2}^o+\frac{1}{2}\ket{S}\ket{\ua},\qquad 
\kettt{\Da_R}^o=\ket{\sigma_L}\ket{\Da_R}=\frac{\sqrt 3}2\kett{1/2,-1/2}^o+\frac{1}{2}\ket{S}\ket{\da},\\
\kettt{\Ua_R}^e&=&\ket{\tau_L}\ket{\Ua_R}=\frac{\sqrt 3}2\kett{1/2,1/2}^e+\frac{1}{2}\ket{S}\ket{2},\qquad
\kettt{\Da_R}^e=\ket{\tau_L}\ket{\Da_R}=\frac{\sqrt 3}2\kett{1/2,-1/2}^e+\frac{1}{2}\ket{S}\ket{0}.
\eea
And for $J_L=0$, the state is
\bea
&&\kettt{\Ua_L}^o=\ket{\Ua_L}\ket{\sigma_R}=\frac{\sqrt 3}2\kett{1/2,1/2}^o-\frac{1}{2}\ket{S}\ket{\ua},\qquad
\kettt{\Da_L}^o=\ket{\Da_L}\ket{\sigma_R}=\frac{\sqrt 3}2\kett{1/2,-1/2}^o-\frac{1}{2}\ket{S}\ket{\da},\\
&&\kettt{\Ua_L}^e=\ket{\Ua_L}\ket{\tau_R}=\frac{\sqrt 3}2\kett{1/2,1/2}^e-\frac{1}{2}\ket{S}\ket{2},\qquad
\kettt{\Da_L}^e=\ket{\Da_L}\ket{\tau_R}=\frac{\sqrt 3}2\kett{1/2,-1/2}^e-\frac{1}{2}\ket{S}\ket{0}.
\eea
To find the effect of $\gamma_{L/R}$ on these we write
\be
3\sqrt{2}\gamma_{L/R}=(L^0\pm M^0)(c\dg_\da+c\dn_\da)-[(L^+\pm M^+)c\dn_\ua+(L^-\pm M^-)c\dg_\ua].
\ee
Acting on $\ket{S}$ term only $M$ yield non-zero result. Therefore, the effect of $3\sqrt{2}\gamma_L$ and $-3\sqrt{2}\gamma_R$ are the same:
\bea
&\ket{S\ua}\to-\ket{T_0}\ket{2}+\sqrt{ 2}\ket{T_+}\ket{0}=\sqrt{3}\kett{1/2,1/2}^e, \qquad
\ket{S\da}\to\ket{T_0}\ket{0}-\sqrt{ 2}\ket{T_-}\ket{2}=\sqrt{3}\kett{1/2,-1/2}^e,\\
&\ket{S2}\to-\ket{T_0}\ket{\ua}+\sqrt{2}\ket{T_+}\ket{\da}=\sqrt{3}\kett{1/2,1/2}^o,\qquad
\ket{S0}\to\ket{T_0}\ket{\da}-\sqrt{ 2}\ket{T_-}\ket{\ua}=\sqrt{3}\kett{1/2,-1/2}^o.
\eea
For the $\kett{1/2,\pm 1/2}^{e/o}$ states we have
\bea
&&3\sqrt{2}\gamma_L\frac{\sqrt{3}}2\kett{1/2,\pm 1/2}^o\to \sqrt{3}\kett{1/2,\pm1/2}^e+\frac{3}{2}\ket{S}\ket{2/0},
\\
&&3\sqrt{2}\gamma_R\frac{\sqrt{3}}2\kett{1/2,\pm 1/2}^o\to \sqrt{3}\kett{1/2,\pm 1/2}^e-\frac{3}{2}\ket{S}\ket{2/0},\\
&&3\sqrt{2}\gamma_L\frac{\sqrt{3}}2\kett{1/2,\pm 1/2}^e\to \sqrt{3}\kett{1/2,\pm1/2}^o+\frac{3}{2}\ket{S}\ket{\ua/\da},
\\
&&3\sqrt{2}\gamma_R\frac{\sqrt{3}}2\kett{1/2,\pm 1/2}^e\to \sqrt{3}\kett{1/2,\pm 1/2}^o-\frac{3}{2}\ket{S}\ket{\ua/\da}.
\eea
Here $\ket{2/0}$ refers to byproducts of $\ket{2}$ for $\kett{1/2,+1/2}^o$ and $\ket{0}$ for $\kett{1/2,-1/2}^o$, respectively. 
So, for $\mu,\nu=\pm 1/2$ one can work out
\be
{}^e\braaa{\nu}\gamma_L\kettt{\mu}^o=\frac{w_L(\alpha)}{\sqrt{2}}\delta_{\mu\nu}, \qquad 
{}^e\braaa{\nu}\gamma_R\kettt{\mu}^o=\frac{w_R(\alpha)}{\sqrt 2}\delta_{\mu\nu},\quad \text{with}\quad
w_{L/R}(\alpha)=\frac{2\cos^2\alpha}{3}[1\mp\sqrt{3}\tan\alpha].
\ee
The gauge connection $A_\alpha^{\mu\nu}=\braaa{\mu}\partial_\alpha\kettt{\nu}=0$ with $\mu,\nu=\Ua^e,\Da^e,\Ua^o,\Da^o$ vanishes. Furthermore, there is no geometric mixing (non-abelian rotation) between $\pm 1/2$ states. Hence the adiabatic protocol produces only dynamical phases within the protected doublet sector.
\subsubsection{Teleportation}
Now, suppose we start with $J_R=0$ and entangle the $\gamma_L$ MZM by coupling it to an auxiliary MZM
\be
H\to H+\Delta H(t), \qquad \Delta H(t)=m(t)P_{L\eta}, \qquad  P_{L\eta}=i\gamma_L\eta', \qquad P_{R\eta}=i\gamma_R\eta'.
\ee
This construction requires an auxiliary fermionic qubit $\sqrt{2}f\dg=\eta'+i\eta''$
\be
\Delta H: \qqquad
\ket{\Updownarrow_R}^o\ket{1}_f\lr -i\ket{\Updownarrow_R}^e\ket{0}_f, \qquad \ket{\Updownarrow_R}_o\ket{0}_f\lr -i\ket{\Updownarrow_R}^e\ket{1}_f.
\ee
So there are two sectors, total even parity and total odd parity. Let us consider the total even-parity sector. Then
\be
\ket{\pm;\Updownarrow_R}^E=\frac{\ket{\Updownarrow_R}^o\ket{1}_f\mp i\ket{\Updownarrow_R}^e\ket{0}_f}{\sqrt{2}}, \qquad \braket{\pm;\Updownarrow_R\vert P_{L\eta}\vert\pm;\Updownarrow_R}=\pm 1/2.
\ee
Suppose we initialize $\psi(t=0)=\ket{-;\Updownarrow_R}^E$ and then turn off the initialization $m(t)\to0$. Then, adiabatically evolve $\alpha(t):-\pi/6\to \pi/6$ by gradually turning on $J_R(t)$ and turning off $J_L(t)$. This corresponds to going from the S=c-S setup to the S-c=S setup. Finally, we measure $P_{R\eta}$. Representing this process by the unitary transformation $U$ we have
\be
U=\exp(-\int{dtH(t)}) \so U\ket{-\Uparrow_R}^E=\ket{-\Uparrow_L}^E.
\ee
In the Heisenberg picture, when projected to the GS manifold
\be
P_G=\sum_{\mu=e/o}\Big(\kettt{\Ua}^{\mu}_\alpha {}^{\mu}_\alpha \braaa{\Ua}+\kettt{\Da}^{\mu}_\alpha {}^{\mu}_\alpha \braaa{\Da}\Big).
\ee
we find
\be
P_G\Big\{U\dg\gamma_L U=\gamma_R\Big\}P_G, \qquad P_G\Big\{U\dg \vec S_R U=\vec S_L\Big\}P_G.
\ee
To see this, note that at the two end points (before and after time evolution) within the GS manifold
\be
\alpha=-\pi/6:\quad \gamma_L\sim \frac{w_L(\alpha)}{\sqrt 2}\dul\tau^x\otimes\bb 1, \quad \vec S_R\sim \bb 1 \otimes\frac{1}{2}\vec{\dul\sigma},\qquad
\alpha=+\pi/6:\quad \gamma_R\sim \frac{w_R(\alpha)}{\sqrt 2}\dul\tau^x\otimes\bb 1, \quad \vec S_L\sim \bb 1 \otimes\frac{1}{2}\vec{\dul\sigma}.
\ee
\subsubsection{Comparison with non-interacting MZMs and the Kondo non-linearity}
Thus, the protocol realizes perfect Majorana teleportation within the projected ground-state manifold. At the end of process $\eta$ is entangled with $\gamma_R$ and the spin-state $\Updownarrow_R$ of the right spin is transferred to the left spin. This agrees with numerics, but they also show that Kondo MZM teleportation is more robust than that of a non-interacting model
\be
H(t)=i[J_L(t)\gamma_L+J_R(t)\gamma_R]\eta',
\ee
for free MZMs $\gamma_{L/R}$. In the latter case, we have
\be
H(t)=i\sqrt{J_L^2+J_R^2}\gamma'(t)\eta',
\ee
with
\be
 \mat{\gamma' \\ \gamma''}=\mat{\cos\theta & \sin\theta \\ -\sin\theta & \cos\theta}\mat{\gamma_L\\\gamma_R},
\ee
where $\tan\theta=J_R/J_L$. So, during the time-evolution $\theta$ evolves from $0$ to $\pi/2$, and therefore $\gamma_L\to \gamma_R$. Therefore, the weights of the two MZMs evolve as $\tan\theta\sim J_R/J_L$ which is linear. 

In contrast, the Kondo anyon case has 
\be
\tan2\alpha=\sqrt{3}\frac{\tan\theta-1}{\tan\theta+1},
\ee
 and the coefficient of the two MZMs is
 \be
 \frac{1-\sqrt{3}\tan\alpha}{1+\sqrt{3}\tan\alpha},
 \ee
  which non-linearly depends on the tunable $\tan\theta$. Thus, the interacting Kondo realization does not reduce to a simple single-particle Majorana interpolation.

By contrast, the free-MZM protocol depends directly on having well-isolated microscopic zero modes and on suppressing all unwanted splittings, overlaps, quasiparticle poisoning, and extra couplings. Small local perturbations can directly deform the zero-mode Hamiltonian. Furthermore, in the free-Majorana protocol, the fermionic zero-mode information only moves downstream: $\gamma_L\to\gamma_R$. There is no accompanying upstream physical spin transport, because the zero mode does not carry an unscreened spin-1/2 degree of freedom.

\subsection{The S-c-c-c-S setup}
Here, we have $c_L$ and $c_R$ and the Hamiltonian is
\be
H=J_L\vec S_L\cdot {\cal J}_L+J_R\vec S_R\cdot {\cal J}_R+[c_0\dg(t_Lc_L+t_Rc_R)+h.c.],
\ee
where ${\cal J}_L$ and ${\cal J}_R$ are defined in terms of $c_L$ and $c_R$ as usual.
The point is that defining independent fermions
\be
\mat{b \\ d}=\frac{1}{\sqrt{\abs{t_L}^2+\abs{t_R}^2}}\mat{t_L & t_R \\ -t_R^* & t_L^*}\mat{c_L \\ c_R},
\ee
the dark (gapless) mode $d$ is decoupled, and only the bright mode $b$ is coupled to $c_0$, gapping each other out, 
\be
H_t=\sqrt{\abs{t_L}^2+\abs{t_R}^2}(c_0\dg b+h.c.)=\sqrt{\abs{t_L}^2+\abs{t_R}^2}(c_+\dg c\dn_+ -c\dg_- c\dn_-),\qquad c_0=\frac{c_++c_-}{\sqrt 2},\quad b=\frac{c_+-c_-}{\sqrt 2}.
\ee
Without loss of generality, we assume the tunneling amplitudes to be real
\be 
\mat{c_L\\ c_R}=\frac{1}{\sqrt{\abs{t_L}^2+\abs{t_R}^2}}\mat{t_L^* & -t_R \\ t_R^* & t_L}\mat{b \\ d}=\mat{\cos\theta & -\sin\theta \\ \sin\theta & \cos\theta}\mat{b \\ d}.
\ee
The Hamiltonian becomes
\be
H=\frac{1}{2}\Big(J_R\vec S_R+J_L\vec S_L\Big)\cdot \Big[\vec{\cal J}_{dd}+\vec{\cal J}_{bb}\Big]+\frac{1}{2}\Big(J_R\vec S_R-J_L\vec S_L\Big)\cdot \Big[\sin2\theta\vec{\cal J}_{(bd)}+\cos2\theta(\vec{\cal J}_{dd}-\vec{\cal J}_{bb})\Big]+\sqrt{t_L^2+t_R^2}(c_0\dg b+h.c.).
\ee
We have used the short-hand notation $\vec{\cal J}_{(bd)}=\vec{\cal J}_{bd}+\vec{\cal J}_{db}$. We can write the Hamiltonian has $H=H_0+V$, where
\be
H_0=\sqrt{t_L^2+t_R^2}[c_0\dg b+h.c.]+(\vec W_1-\cos2\theta\vec W_2)\cdot\vec{\cal J}_{dd},\quad 
\text{with}\quad V=(\vec W_1+\cos2\theta\vec W_2)\cdot \vec{\cal J}_{bb}-\sin2\theta\vec W_2\cdot \vec{\cal J}_{(bd)},\qquad
\ee
\be
2\vec W_1=J_L\vec S_L+J_R\vec S_R=\bar J\vec L+\frac{\delta J}{2}\vec M,\qquad \text{and},\qquad 
2\vec W_2=J_L\vec S_L-J_R\vec S_R=\bar J\vec M+\frac{\delta J}{2}\vec L.
\ee
The low-energy sector (projecting out the gapped bright mode $b$) is
\be
H_0=\sqrt{t_L^2+t_R^2}[c_0\dg b+h.c.]+(J_L\sin^2\theta \vec S_L+J_R\cos^2\theta \vec S_R)\cdot\vec{\cal J}_{dd}.\label{eqS82}
\ee
whose GS we already know from the S-c-S setup.
\bea
\left\{\matn{
\kettt{\Ua}^{o/e}\equiv\cos\alpha\kett{1/2,1/2}^{o/e}-\sin\alpha\ket{S}\ket{\ua/2},\\
\kettt{\Da}^{o/e}\equiv\cos\alpha\kett{1/2,-1/2}^{o/e}-\sin\alpha\ket{S}\ket{\da/0}
}\right\}_{d},\quad \otimes \ket{\ua\da}_{c_-} \otimes \ket{0}_{c_+}.\label{eqS83}
\eea
There, we saw that by a $J_L-J_R$ sweep the Kondo MZM can be teleported with a backflow of the spin. Eq.\,\pref{eqS82} shows that the same transport can also be implemented in the S-c-c-c-S setup with constant $\tan\theta=t_R/t_L$. Moreover, it is also possible to do the same by sweeping $t_L-t_R$ at constant $J_L=J_R$. We will use the latter for braiding later on.
\subsubsection{The leading $J/t$ correction}
To see the effect of coupling to the gapped $b$ fermions, we can use a Schrieffer-Wolff transformation to find the effective energy after integrating out the high energy sector. For simplicity, we assume $t_L=t_R$ or $\theta=\pi/4$ in this section. Defining $P$ as the projector to the low-energy states \pref{eqS83} and $Q=\bb 1-P$, the effective Hamiltonian is given by
\be
H_{\rm eff}=PHP+PVQ\frac{1}{E_0-QH_0Q}QVP.
\ee
We work to order $J/t$. Assuming $t\gg J$, we can assume all excited states have the same energy, and write
\be
H_{\rm eff}=H_0-\frac{1}{t\sqrt 2}PV^2P.
\ee
It is straightforward to show that
\bea
PV^2P&=&\frac{1}{4}\vec W_1^2+\frac{1}{4}[2\vec W_2^2+(W_2^-W_2^+-W_2^+W_2^-)(d\dg_\ua d\dn_\ua-1/2)]+{W_2^z}^2\\
&&\hspace{2cm}+\frac{1}{4}(d\dn_\da+d\dg_\da)d\dg_\ua [W_2^-,W_2^z]+\frac{1}{4}(d\dn_\da+d\dg_\da)d\dn_\ua [W_2^+,W_2^z].
\eea
There will be no $W_1W_2$ cross term in the projected space. We find that the degenerate GS manifold,
\bea
&&\kettt{\Ua}^{o/e}\equiv\cos\alpha \Big[\sqrt{2/3}\ket{T_+}\ket{\da/0}-\sqrt{1/3}\ket{T_0}\ket{\ua/2}\Big]-\sin\alpha\ket{S}\ket{\ua/2},
\\
&&\kettt{\Da}^{o/e}\equiv\cos\alpha\Big[\sqrt{1/3}\ket{T_0}\ket{\da/0}-\sqrt{2/3}\ket{T_-}\ket{\ua/2}\Big]-\sin\alpha\ket{S}\ket{\da/0},
\eea
remains degenerate under $H_{\rm eff}$ and only the energy is shifted from $E_0=-\frac{1}{2}\bar J-\sqrt{(\bar J/2)^2+\frac{3}{16}\delta J^2}$ to
\be
E
=
E_0-\frac{1}{t\sqrt{2}}
\Big[
\frac{13}{4}\bar J^2-\frac{3}{4}J_LJ_R
+
\Big(\bar J^2-\frac{5}{3}J_LJ_R\Big)\cos^2\alpha
-\frac{\sqrt{3}}{4}(J_L^2-J_R^2)\sin\alpha\cos\alpha
\Big].
\ee
This is a consequence of the conservation of fermion parity and the SU(2) symmetry; the only effect of the gapped mode is to change the U(1) dynamic phase accumulated during the adiabatic evolution.
\subsection{The c-S-c setup}
Next, we move to the c-S-c setup, with two sets of currents ${\cal J}^L$ and ${\cal J}^R$ and the Hamiltonian
\be
H=-i\vec S\cdot\big(J_L\vec{\cal J}_L+J_R\vec{\cal J}_R\big), \qquad {\cal J}_a^{R/L}=i\eps^{abc}\chi^\mu_b\chi^\mu_c, \qquad \mu=R,L.
\ee
This analysis parallels the previous construction, with analogous definitions for $\vec L$ and $\vec M$. 
We have ${\cal J}_L^2={\cal J}_R^2=3/4$. Expressing
\be
\vec{\cal J}^{R/L}=\frac{1}{2}(\vec L\pm \vec M) \so [{\cal J}_a^R,{\cal J}^R_b]=i\eps_{abc}{\cal J}^R_c, \quad [{\cal J}_a^L,{\cal J}^L_b]=i\eps_{abc}{\cal J}^L_c, \quad [{\cal J}_a^R,{\cal J}^L_b]=0,
\ee
the Kondo interaction is
\bea
H&=&\vec S\cdot\Big[\bar J\vec L+\frac{1}{2}\delta J\vec M\Big]\qqquad \bar J\equiv\frac{J_L+J_R}2,\quad \delta J \equiv J_L-J_R.
\eea
We define $\sqrt{2}c=\chi^L-i\chi^R$, using which we can define a set of currents:
\be
L_x=-i(c\dg_yc_z-c\dg_zc_y), \qquad L_y=-i(c\dg_zc_x-c\dg_xc_z), \qquad L_z=-i(c\dg_xc_y-c\dg_yc_x),
\ee
obeying the SU(2) algebra $[L_x,L_y]=iL_z$, $[L_y,L_z]=iL_x$ and $[L_z,L_x]=iL_y$, and having the Casimir
\be
\vec L^2=N(3-N)=\ell(\ell+1) \so \ell(N=0)=\ell(N=3)=0, \qquad \ell(N=1)=\ell(N=2)=1.
\ee
We can also define a new set of currents:
\be
M_x=-i(c\dg_yc\dg_z-c_zc_y), \qquad M_y=-i(c\dg_zc\dg_x-c_xc_z), \qquad M_z=-i(c\dg_xc\dg_y-c_yc_x),
\ee
which obey a more complicated algebra (but note that $N$ and thus $\vec M^2$ do not commute with $\vec M$)
\be
[M_x,M_y]=iL_z, \qquad [M_y,M_z]=iL_x, \qquad [M_z,M_x]=iL_y, \qquad \vec M^2=N(N-3)+3.
\ee
We also have $[L_a,M_b]=i\eps_{abc}M_c$. If $J_L=J_R$ we can write
\be
H=\frac{{\bar J}}{2}[j(j+1)-3/4-\ell(\ell+1)]={\bar J}\times\left\{\matl{ 1/2 & \ell=1, j=3/2 \\ 0\qquad & \ell=0 \\ -1 & \ell=1, j=1/2}\right.
,
\ee
where we have used that $1_3\otimes ({1}/{2})_2=(1/2)_2+(3/2)_4$. The eigenstates $\kett{j,m_j}$ in terms of $\ket{\ell,m_\ell}$, with the right Clebsch-Gordan coefficients, are
\be
\kett{1/2,1/2}\equiv\sqrt{2/3}\ket{1,1}\ket{\Da}-\sqrt{1/3}\ket{1,0}\ket{\Ua}, \qquad \kett{1/2,-1/2}\equiv\sqrt{1/3}\ket{1,0}\ket{\Da}-\sqrt{2/3}\ket{1,-1}\ket{\Ua}.
\ee
Note that acting with $(L^-+S^-)$ on $\kett{1/2}$ gives $\kett{-1/2}$, using $L^{-1}\ket{1,1}=\sqrt{2}\ket{1,0}$ and $L^-\ket{1,0}=\sqrt{2}\ket{1,-1}$. Using $L_z$ to label $m_\ell$ and the raising operator $L^+=L_x+iL_y$ obeying $[L_z,L^+]=L^+$, they are expressed as
\be
L^+=\sqrt{2}(c\dg_0 c\dn_{-1}+c\dg_1 c\dn_0),\qqquad L_z=c\dg_1c\dn_1-c\dg_{-1}c\dn_{-1},\qquad [L_z,c\dg_m]=mc\dg_m, \quad [L_z,c\dn_m]=-mc\dn_m,
\ee
in terms of transformed fermions
\be
c_0=c_z, \andd c\dg_1=-\frac{c\dg_x+ic\dg_y}{\sqrt 2}, \qquad c\dg_{-1}=\frac{c\dg_x-ic\dg_y}{\sqrt 2}.
\ee
For $N=1$, we have $\ket{\ell=1,m}_{N=1}=c\dg_m\ket{0}$. For $N=2$ we define $\ket{\Omega}=ic\dg_xc\dg_yc\dg_z\ket{0}=c\dg_1 c\dg_{-1}c\dg_0\ket{0}$. The $L_z$ operator does not fix the sign of all the states. To fix the overall phase, we make an assumption for $\ket{1,1}_2$ and apply $L^-$ operator:
\be
\ket{1,1}_{N=2}=-c_{-1}\ket{\Omega}=c\dg_1c\dg_0\ket{0}, \qquad \ket{1,0}_{N=2}=c_0\ket{\Omega}=c\dg_1c\dg_{-1}\ket{0}, \quad \ket{1,-1}_{N=2}=-c_1\ket{\Omega}=-c\dg_{-1}c\dg_0\ket{0}.
\ee
\subsubsection{Kondo coupling anisotropy}
For $\delta J=0$, the GS has a fourfold degeneracy, ${\cal H}=(1/2)_2^{even}\oplus(1/2)_2^{odd}$:
\be
\ket{GS_1}=\kett{1/2}_{N=2}, \quad \ket{GS_2}=\kett{-1/2}_{N=2}, \quad \ket{GS_3}=\kett{1/2}_{N=1}, \quad
\ket{GS_4}=\kett{-1/2}_{N=1}, 
\ee
For $\delta J\neq 0$, the $\vec M$ sector couples the $\ell=1, j=1/2$ GS quartet with the $\ell=0,j=1/2$ excited state quartet. The parity is conserved. 
In $\ell$ language $\vec M$ operator is
\be
M^+=\sqrt{2}(c\dg_1c\dg_0+c_0c_{-1}),\quad M_z=-(c\dg_1c\dg_{-1}+c_{-1}c_1).
\ee
Thus, the action of
\be
\hat V/(\delta J/2)=\frac{1}{2}[M^+S^-+M^-S^+]+M^zS^z,
\ee
on the Nozi\`eres doublet is ($V\equiv {\sqrt 3}\delta J/4$)
\be
\hat V\kett{1/2,1/2}_1=V\ket{3}\ket{\Ua}, \quad
\hat V\kett{1/2,-1/2}_1=V\ket{3}\ket{\Da},\qquad
\hat V\kett{1/2,1/2}_2=V\ket{0}\ket{\Ua},\quad
\hat V\kett{1/2,-1/2}_2=V\ket{0}\ket{\Da}.
\ee
The effective 2x2 coupling is
\be
H=\bar J\mat{-1 \\ & 0}+\mat{& V \\ V &}.
\ee
The diagonalization is done using the unitary transformation $U$
\be
  U=e^{i{\alpha}\dul\sigma^y}=\cos\alpha\bb 1+i\dul\sigma^y\sin\alpha, \qquad
U\dg HU=-\frac{1}{2}\bar J\bb 1-\dul\sigma^z[\frac{1}{2}\bar J\cos2\alpha+V\sin2\alpha]+\dul\sigma^x[-\bar J\sin2\alpha+V\cos2\alpha].
\ee
Choosing $\tan2\alpha={2V}/{\bar J}$, the Hamiltonian is diagonalized and we find that there is a level-repulsion
\be
 E_\pm=-(\bar J/2)\pm \sqrt{(\bar J/2)^2+V^2}=-\frac{1}{2}\bar J\pm \sqrt{(\bar J/2)^2+\frac{3}{16}\delta J^2}.
\ee
So, we find that the GS is $U\dg\mat{1\\0}$ in the odd/even parity sectors, and is given by
\be
\kettt{\Ua/\Da}_\alpha^{o}\equiv\cos\alpha\kett{\pm 1/2,1/2}_1-\sin\alpha\ket{3}\ket{\Ua/ \Da},\qquad
\kettt{\Ua/\Da}_\alpha^{e}\equiv\cos\alpha\kett{\pm 1/2,1/2}_2-\sin\alpha\ket{0}\ket{\Ua/ \Da}.
\ee
Note that $\kett{1/2,\pm 1/2}_1$ has one electron and $\kett{1/2,\pm 1/2}_2$ has two electrons, and they are mixed with $\ket{3}$ with three electrons and $\ket{0}$ with zero electrons and appropriate spins, respectively. For $J_R=0$ we have $\delta J=2\bar J$ and therefore
\be
J_R=0\so \tan2\alpha_*=\sqrt{3} \so \alpha_*=\pi/6 \so \cos\alpha_*=\sqrt{3}/2,\quad \sin\alpha_*=1/2.
\label{eq42}
\ee
Forming the gauge connection $A_\alpha^{\mu\nu}=\braaa{\mu}\partial_\alpha\kettt{\nu}=0$ with $\mu,\nu=\pm1/2$, we see that it vanishes. Therefore, there is no geometric mixing (non-abelian rotation) between $\pm 1/2$ states.
\subsubsection{Matrix elements of $\gamma_L$ and $\gamma_R$}
Let's see how $\gamma_L=\frac{2}{3}\vec S\cdot\vec\chi_L$ and $\gamma_R=\frac{2}{3}\vec S\cdot\vec\chi_R$ act in this space. Using
\bea
3\sqrt{2}\gamma_L&=&2S^z(c\dg_0+c\dn_0)+\sqrt{2}[S^+(c\dg_{-1}-c\dn_1)-S^-(c\dg_1-c\dn_{-1})], \\
3\sqrt{2}i\gamma_R&=&2S^z(c\dg_0-c\dn_0)+\sqrt{2}[S^+(c\dg_{-1}+c\dn_1)-S^-(c\dg_1+c\dn_{-1})],
\eea
we find that the action of $\gamma_L$ closes within the subset
\bea
\matl{\kettt{\Ua}^{o}_R\propto \frac{\sqrt{3}}{2}\kett{1/2,1/2}_1-\frac{1}{2}\ket{3}\ket{\Ua}\\ 
\kettt{\Da}^{o}_R\propto \frac{\sqrt{3}}{2}\kett{1/2,-1/2}_1-\frac{1}{2}\ket{3}\ket{\Da}}, \andd
\matl{ 
\kettt{\Ua}_R^e\propto \frac{\sqrt{3}}{2}\kett{1/2,1/2}_2-\frac{1}{2}\ket{0}\ket{\Ua}\\ 
\kettt{\Da}_R^e\propto \frac{\sqrt{3}}{2}\kett{1/2,-1/2}_2-\frac{1}{2}\ket{0}\ket{\Da}}.
\eea
Consequently:
\be
\sqrt{2}\gamma_L:\quad \kettt{\Ua}_R^e\to\kettt{\Ua}_R^o,\quad \kettt{\Ua}_R^o\to\kettt{\Ua}_R^e \andd
\kettt{\Da}_R^e\to\kettt{\Da}_R^o,\quad \kettt{\Da}_R^o\to\kettt{\Da}_R^e.
\ee
Likewise, we have
\be
\sqrt{2}i\gamma_R:\quad \kettt{\Ua}_L^e\to\kettt{\Ua}_L^o,\quad \kettt{\Ua}_L^o\to-\kettt{\Ua}_L^e \andd
\kettt{\Da}_L^e\to\kettt{\Da}_L^o,\quad \kettt{\Da}_L^o\to-\kettt{\Da}_L^e.
\ee
The minus sign is needed because the operator on the left squares to $-1$. For more general $\alpha$ we have
\be
3\matn{\sqrt{2}\gamma_L \\ \sqrt{2}i\gamma_R}\kettt{\Updownarrow}_\alpha^o=(2\cos\alpha\pm\sqrt{3}\sin\alpha)\kett{1/2}_2\mp\sqrt{3}\cos\alpha\ket{0}\ket{\Ua}=3\matn{w_L(\alpha) \\ w_R(\alpha)}\kettt{\Updownarrow}_\alpha^e+\dots,
\ee
and
\be
3\matn{\sqrt{2}\gamma_L \\ \sqrt{2}i\gamma_R}\kettt{\Updownarrow}_\alpha^e=(\pm 2\cos\alpha+\sqrt{3}\sin\alpha)\kett{1/2}_2-\sqrt{3}\cos\alpha\ket{0}\ket{\Ua}=3\matn{w_L(\alpha) \\ w_R(\alpha)}\kettt{\Updownarrow}_\alpha^e+\dots,
\ee
where the $\dots$ refer to excited states of the $\hat V$ hybridization. So, for $\mu,\nu=\pm 1/2$ one finds
\be
{}_\alpha^e\braaa{\nu}\gamma_L\kettt{\mu}_\alpha^o=\frac{w_L(\alpha)}{\sqrt{2}}\delta_{\mu\nu}, \qquad 
{}_\alpha^e\braaa{\nu}i\gamma_R\kettt{\mu}_\alpha^o=\frac{w_R(\alpha)}{\sqrt 2}\delta_{\mu\nu},\qquad
w_{L/R}(\alpha)=\frac{2\cos^2\alpha}{3}[1\pm\sqrt{3}\tan\alpha].
\ee
This can be verified by explicitly writing the end-point wavefunctions in terms of the $\sigma-\tau$ wavefunctions. This amounts to a mapping between the economical dense coding of $\kettt{\Updownarrow}^{o/e}$ and the sparse $\sigma-\tau$ coding of the 2CK site.
\subsubsection{Mapping between dense and sparse codings at $J_R=0$}
In the above discussion, the wavefunction was written without any reference to $\chi^0$, which was decoupled. Although ${\cal J}_L$ and ${\cal J}_R$ are each $Z_2$ anomalous, together the Hilbert space was anomaly free, with the Hilbert space
\be
e_\ua:\,\, \Big\{{\vert{\frac12,\frac12}\rangle\rr\rangle}_2,\ket{\Ua}\ket{0}\Big\},\quad
e_\da:\,\, \Big\{{\vert{\frac12,-\frac12}\rangle\rr\rangle}_2,\ket{\Da}\ket{0}\Big\}
\quad
o_\ua:\,\, \Big\{{\vert{\frac12,\frac12}\rangle\rr\rangle}_1,\ket{\Ua}\ket{3}\Big\},\quad
o_\da:\,\, \Big\{{\vert{\frac12,-\frac12}\rangle\rr\rangle}_1,\ket{\Da}\ket{3}\Big\}.
\ee
However, the four GSs live in a subset of these states, made out of a particular $\alpha$-dependent linear combination within each sector. In the following we map this dense coding to the sparse coding which uses the local Hilbert space, including the decoupled $\chi^0$. For $J_R=0$ in the c=S-c setup the GS is one of the following (depending on parity):
\be
e_\ua:\, \Big\{\ket{\sigma_L}\ket{\ua_R},\ket{\tau_L}\ket{2_R}\Big\},\quad 
e_\da:\, \Big\{\ket{\sigma_L}\ket{\da_R},\ket{\tau_L}\ket{0_R}\Big\},\quad
o_\ua:\, \Big\{\ket{\tau_L}\ket{\ua_R},\ket{\sigma_L}\ket{2_R}\Big\},\quad
o_\da:\, \Big\{\ket{\sigma_L}\ket{0_R},\ket{\tau_L}\ket{\da_R}\Big\},\label{eq43}
\ee
where we have represented the Kondo singlet with a left-fermion in spin and isospin sectors with $\ket{\sigma_L}$ and $\ket{\tau_L}$, respectively. To connect this sparse basis to the dense basis used before (without the $\chi^0$ basis), we introduce a $Z_2$ gauge symmetry $\chi^0_{R/L}\to-\chi^0_{R/L}$, with the constraint $:\hspace{-.075cm}d\dg d\hspace{-.075cm}:\equiv i\chi_R^{0}\chi_L^{0}$. 

To apply the $d$-constraint we can form linear combinations with good $d\dg d$ out of \pref{eq43}. It can be shown, after a tedious but straightforward algebra, that the states involving the odd dense sector (first two lines) and even dense sector (last two lines) are
\bea
&&\frac{1}{\sqrt{2}}\mat{\ket{\sigma_L2_R-i\tau_L\rr\ua_R}\\ \ket{\sigma_L\rr\ua_R-i\tau_L2_R}}=\Big\{\frac{\sqrt3}{2}\kett{1/2,+1/2}_1-\frac{1}{ 2}\ket{\Ua}\ket{3}\Big\}\otimes\mat{ \ket{0}_d \\ \ket{1}_d}=\kettt{\Ua}_R^o\otimes\mat{ \ket{0}_d \\ \ket{1}_d} \\
&&\frac{1}{\sqrt 2}\mat{\ket{\sigma_L0_R-i\tau_L\rr\da_R} \\ \ket{\sigma_L\rr\da_R-i\tau_L0_R}}=\Big\{\frac{\sqrt{3}}{2}\kett{1/2,-1/2}_1-\frac{1}{2}\ket{\Da}\ket{3}\Big\}\otimes\mat{\ket{0}_d \\ \ket{1}_d}=\kettt{\Da}_R^o\otimes\mat{\ket{0}_d \\ \ket{1}_d},\\
&&\frac{1}{\sqrt{2}}\mat{\ket{\tau_L2_R-i\sigma_L\rr\ua_R}\\ \ket{\tau_L\rr\ua_R-i\sigma_L2_R}}=\Big\{\frac{\sqrt3}{2}\kett{1/2,1/2}_2-\frac{1}{ 2}\ket{\Ua}\ket{0}\Big\}\otimes\mat{ \ket{0}_d \\ \ket{1}_d}=\kettt{\Ua}_R^e\otimes\mat{ \ket{0}_d \\ \ket{1}_d},\\
&&\frac{1}{\sqrt 2}\mat{\ket{\tau_L0_R-i\sigma_L\da_R} \\ \ket{\tau_L\da_R-i\sigma_L0_R}}=\Big\{\frac{\sqrt{3}}{2}\kett{1/2,-1/2}_2-\frac{1}{2}\ket{\Da}\ket{0}\Big\}\otimes\mat{\ket{0}_d \\ \ket{1}_d}=\kettt{\Da}_R^e\otimes\mat{ \ket{0}_d \\ \ket{1}_d}.
\eea
In each line, the first (second) element of the $d$-spinor corresponds to having empty (filled) $d$ electron states. 

We can apply the gauge constraint and only keep one sector. Assume the $\vec{\dul\tau}$ and $\vec{\dul\sigma}$ matrices act in the (dense) o/e and $\{\ket{0}_d,\ket{1}_d\}$, respectively. Then, the effect of $\gamma_L$, $\chi^0_L$, and $\sqrt{2}\chi^0_R=-i(c\dg_{R\da}-c\dn_{R\da})$ at the $J_R=0$ end point is
\be
\Psi=\mat{\kettt{\Updownarrow}_R^o \\ \kettt{\Updownarrow}_R^e}\otimes\mat{\ket{0}_d \\ \ket{1}_d},\qquad
\gamma_L\,\Psi=\frac{1}{\sqrt 2}\dul\tau^x\,\Psi.
\ee
\subsubsection{Mapping between dense and sparse codings at $J_L=0$}
In this case, the anomalous basis is now obtained by interchanging $L\leftrightarrow R$ in \pref{eq43}, namely
\be
e_\ua:\, \Big\{\ket{\ua_L}\ket{\sigma_R},\ket{2_L}\ket{\tau_R}\Big\},\quad
e_\da:\,\Big\{\ket{\da_L}\ket{\sigma_R},\ket{0_L}\ket{\tau_R}\Big\},\quad
o_\ua:\, \Big\{\ket{2_L}\ket{\sigma_R},\ket{\ua_L}\ket{\tau_R}\Big\},\quad
o_\da:\,\Big\{\ket{0_L}\ket{\sigma_R},\ket{\da_L}\ket{\tau_R}\Big\}.
\ee
The states involving the odd and even dense sectors are
\bea
&&\frac{1}{\sqrt2}\mat{\ket{\ua_L\rr\tau_R-i2_L\sigma_R}\\ \ket{\ua_L\rr\sigma_R-i2_L\tau_R}}
=
\Big\{\frac{\sqrt3}{2}\kett{1/2,1/2}_1+\frac{1}{2}\ket{3}\ket{\Ua}\Big\}\otimes\mat{\ket{0}_d \\ \ket{1}_d}
=\kettt{\Ua}_L^o\otimes\mat{\ket{0}_d \\ \ket{1}_d},\\
&&\frac{1}{\sqrt2}\mat{\ket{\da_L\rr\tau_R-i0_L\sigma_R}\\ \ket{\da_L\rr\sigma_R-i0_L\tau_R}}
=
\Big\{\frac{\sqrt3}{2}\kett{1/2,-1/2}_1+\frac{1}{2}\ket{3}\ket{\Da}\Big\}\otimes\mat{\ket{0}_d \\ \ket{1}_d}
=\kettt{\Da}_L^o\otimes\mat{\ket{0}_d \\ \ket{1}_d},\\
&&\frac{1}{\sqrt2}\mat{\ket{2_L\tau_R-i\rr\ua_L\rr\sigma_R}\\ \ket{2_L\sigma_R-i\rr\ua_L\rr\tau_R}}
=
\Big\{\frac{\sqrt3}{2}\kett{1/2,1/2}_2+\frac{1}{2}\ket{0}\ket{\Ua}\Big\}\otimes\mat{\ket{0}_d \\ \ket{1}_d}
=\kettt{\Ua}_L^e\otimes\mat{\ket{0}_d \\ \ket{1}_d},\\
&&\frac{1}{\sqrt2}\mat{\ket{0_L\tau_R-i\rr\da_L\rr\sigma_R}\\ \ket{0_L\sigma_R-i\rr\da_L\rr\tau_R}}
=
\Big\{\frac{\sqrt3}{2}\kett{1/2,-1/2}_2+\frac{1}{2}\ket{0}\ket{\Da}\Big\}\otimes\mat{\ket{0}_d \\ \ket{1}_d}
=\kettt{\Da}_L^e\otimes\mat{\ket{0}_d \\ \ket{1}_d}.
\eea
We can apply the gauge constraint and only keep one sector. Then for the action of $\gamma_R$, $\chi^0_R$ and $\chi^0_L=-i(c\dg_{L\da}-c\dn_{L\da})$ at the $J_L=0$ end point we have
\be
\Psi=\mat{\kettt{\Updownarrow}_L^o \\ \kettt{\Updownarrow}_L^e}\otimes\mat{\ket{0}_d \\ \ket{1}_d},\qquad
\gamma_R\,\Psi=\frac{1}{\sqrt 2}\dul\tau^y\,\Psi.
\ee
\subsubsection{Failed switching across spin}
Let us start with $w_{L,R}\neq 0$ (both non-zero but $w_R\sim 0$). Then, we entangle $\gamma_L$ and $\gamma_R$ with auxiliary MZMs $\eta'$ and $\eta''$, respectively:
\be
H\to H+\Delta H(t), \qquad \Delta H(t)=m_L(t)P_{L\eta'}+m_R(t)P_{R\eta''}, \qquad  P_{L\eta'}=i\gamma_L\eta', \qquad P_{R\eta''}=i\gamma_R\eta''.
\ee
This clearly needs an additional qubit $\sqrt{2}f\dg=\eta'+i\eta''$ (an economic version would be to use $\eta=\chi^0=\chi''_\da$, but that would couple $\kettt{\Ua}_\alpha^e$ to $\kettt{\Da}_\alpha^o$ and so on). So in the space of 
\be
\Big\{\kettt{\Updownarrow}_\alpha^o\ket{1}_f,\kettt{\Updownarrow}_\alpha^e\ket{0}_f\Big\}^E\oplus\Big\{\kettt{\Updownarrow}_\alpha^e\ket{1}_f,\kettt{\Updownarrow}_\alpha^o\ket{0}_f\Big\}^O.
\ee
Notice that individual $\gamma_L$ or $\eta'$ would cause transitions between these E and O sectors; but their product acts within E or O sector and thus can be represented with $\sigma^{x,y}$ or possibly $\sigma^{x,y}\tau^z$ alone. So, once $\gamma_L$ and $\eta'$ are entangled, simultaneously $\gamma_R$ and $\eta''$ are also entangled, which might seem counter intuitive, but is reasonable considering that we are assuming the total parity is conserved.

Without loss of generality, we focus on the even parity sector. In this space the operators have the representations
\be
{\rm Even}:\quad P_{L\eta'}=\frac{w_L}{2}\dul\tau^y, \qquad P_{L\eta''}=\frac{w_L}{2}\dul\tau^x, \qqquad 
P_{R\eta'}=-\frac{w_R}{2}\dul\tau^x, \quad P_{R\eta''}=\frac{w_R}{2}\dul\tau^y.
\ee
The eigenstates and eigenvalues of $ H$ are therefore,
\be
\ket{\pm,\Updownarrow}_\alpha^E\equiv\frac{1}{\sqrt 2}[\kettt{\Updownarrow}_\alpha^e\ket{0}_f\pm i \kettt{\Updownarrow}_\alpha^o\ket{1}_f],\qquad E_{\pm,E}=\pm \frac{1}{2}(w_Rm_R+w_Lm_L).
\ee
So, accounting for the spin ($\Ua/\Da$) and total parity ($E/O$), the GS is four-fold degenerate. The choice of the initial state depends on history.  Clearly, the spin degree of freedom is an spectator and we can neglect it. With $w_Rm_R+w_Lm_L>0$, assuming we initialize the state in the total-even, spin-up sector
\be
\ket{\psi}_i=\ket{-,\Ua}_{\pi/6}^E.
\ee
Then turn off the initialization $m(t)\to0$, and adiabatically evolve $\alpha(t):+\pi/6\to -\pi/6$ by gradually turning on $J_R(t)$ and turning off $J_L(t)$. This corresponds to going from the c=S-c setup to the c-S=c setup. Finally, we measure $P_{R\eta}$. Representing this process by the unitary transformation $\hat U$, up to an overall dynamical phase we have
\be
\hat U=\exp\Big[-i\int{dtH(t)}\Big]\sim \bb 1\quad \to\quad \hat U\mat{\kettt{\Updownarrow}_R^e \\ \kettt{\Updownarrow}_R^o}\sim \bb 1\mat{\kettt{\Updownarrow}_L^e \\ \kettt{\Updownarrow}_L^o} \so \ket{\psi}_f=\hat U\ket{-,\Ua}_{\pi/6}^E\sim \ket{-,\Ua}_{-\pi/6}^E.
\ee

As we can see, these are $\braket{P_{L\eta'}}\propto\braket{P_{R\eta''}}$ and $\braket{P_{L\eta''}}\propto\braket{P_{R\eta'}}$, and only the proportionality constant depends on $\alpha$. Since the initial state was an eigenstate of $\dul\tau^y$, it remains so and $\sbraket{\dul\tau^x}=0$ remains zero. Therefore, clearly $\alpha(t)=\pi/2\to -\pi/2$ will not help in switching the Kondo MZM across the spin.

\subsubsection{Aadditional tunneling term}
Next, we add a tunneling between the two $c$-s in c-S-c. In the $c$-language, this translates to a population term 
\be
\Delta H_e=\eps_c:\hspace{-.075cm}N_c\hspace{-.075cm}:, \qquad :N_c:=c\dg_1c\dn_1+c\dg_0c\dn_0+c\dg_{-1}c\dn_{-1}-3/2 \,=\,i(\chi^x_L\chi^x_R+\chi^y_L\chi^y_R+\chi^z_L\chi^z_R). 
\ee
This term lifts the degeneracy of the $N_c=1$ and $N_c=3$ states in the odd sector, and the degeneracy of the $N_c=0$ and $N_c=2$ states in the even sector, thereby changing the mixing angle $\alpha$. For the odd/even GS
\be
\Delta H_{o/e}=\pm\eps_c\mat{-1/2 \\ & 3/2},\quad\to\quad 
E_\pm^o=-\frac{1}{2}\bar J\pm\frac{1}{2}\eps_c\pm \sqrt{(\bar J/2\pm \eps_c)^2+\frac{3}{16}\delta J^2}, \qquad \tan2\alpha_{o/e}=\frac{V}{\bar J/2\pm\eps_c},
\ee
where $+\eps_c$ refers to the odd sector and $-\eps_c$ to the even sector. Therefore, $\Delta H$ acts as $\dul\tau^z$ in the space of $\ket{\Updownarrow}^o$ and $\ket{\Updownarrow}^e$, and lifts the degeneracy between even and odd sectors. Expanding to small values of $\eps_c$, we have the even-odd splitting
\be
\Delta H=\eps_c\dul\tau^z\Big(\frac{1}{2}-\cos2\alpha\Big)+O(\eps_c^2).
\ee
Note that the interaction $UN_c^2$ would not split the even-odd sectors, as $(\pm\eps_c)^2=\eps_c^2$ is blind to the fermion parity.

\subsubsection{Successful (non-topological) switching with the tunneling term}
Now if $\eps_c$ is non-zero, even/odd sectors are not degenerate and we have
\be
\hat U=\exp\Big[-i\int{dt\Delta H_c(t)}\Big]\sim \exp\Big[-i\int{dt\eps_c(t)[1/2-\cos2\alpha(t)]\dul\tau^z}\Big],
\ee
so that
\be
\hat U\mat{\kettt{\Updownarrow}_R^e \\ \kettt{\Updownarrow}_R^o}=U\mat{\kettt{\Updownarrow}_L^e \\ \kettt{\Updownarrow}_L^o}, \qquad U=\exp\Big({i\vartheta\dul\tau^z}\Big).
\ee
Since we have
\be
U\dg\dul\tau^xU=\cos2\vartheta\dul\tau^x+\sin2\vartheta\dul\tau^y, \andd 
U\dg\dul\tau^yU=\cos2\vartheta\dul\tau^y-\sin2\vartheta\dul\tau^x,
\ee
the unitary transformations rotates $\dul\tau^y$ to $\dul\tau^x$ and vice versal; whose precession depends on $\alpha$. For the special choice of $\eps_c=\dot\alpha/2A$ the $\hat U$-operation is path independent 
\be
\vartheta=-\int_{t_i}^{t_f}\eps_c(t)[1/2-\cos2\alpha(t)]=\frac{1}{2A}\int_{-\pi/6}^{\pi/6}{d\alpha}[1/2-\cos2\alpha]=\frac{\pi}{4A}(\frac{1}{3}-\frac{\sqrt 3}{\pi}),
\ee
and the special choice of $A=(1/3-\sqrt3/\pi)$ gives $\vartheta=\pi/4$. 
For this particular choice,
\be
P_{R\eta'}(t_f)\propto \bra{\psi(t)} \dul\tau^x\ket{\psi(t)}=\bra{\psi} \dul\tau^y\ket{\psi}\propto P_{L\eta'}(t_i), \qquad
P_{L\eta''}(t_f)\propto \bra{\psi(t)} \dul\tau^x\ket{\psi(t)}=-\bra{\psi} \dul\tau^y\ket{\psi}\propto P_{R\eta''}(t_i).
\ee
Therefore, we have
\be
\gamma_L\to\gamma_R, \andd \gamma_R\to-\gamma_L,
\ee
which can be represented by the unitary transformation
\be
\hat U=\frac{1+2\gamma_L\gamma_R}{\sqrt{2}}.
\ee
\subsection{The S-c-c-S setup}
Next we consider the following problem
\be
H=J_L\vec S_L\cdot\vec{\cal J}_L+J_R\vec S_R\cdot\vec{\cal J}_R+H_t, \quad H_t=-(tc\dg_Lc\dn_R+h.c.)=-i\abs{t}\sum_{\mu=0,x,y,z} \chi^\mu_L\chi^\mu_R.
\ee
In terms of $\ket{\sigma^\mu}$ and $\ket{\tau^\mu}$ with $\mu=0,x,y,z$ defined in Eqs.\,\pref{eqS35} and \pref{eqS36} we have
\be
\braket{\tau^\mu\vert\chi^x\vert\sigma^\nu}=\matc{cc|cc}{0 & 1&0&0 \\ 1 &0&0&0 \\\hline 0&0&0& i \\ 0&0&-i&0}, \quad 
\braket{\tau^\mu\vert\chi^y\vert\sigma^\nu}=\matc{cc|cc}{0 & 0&1&0 \\ 0 &0& 0 & -i \\\hline 1&0&0& 0 \\ 0&i&0&0}, \quad 
\braket{\tau^\mu\vert\chi^z\vert\sigma^\nu}=\matc{cc|cc}{0 & 0&0&1 \\ 0 &0& i & 0 \\\hline 0&-i&0& 0 \\ 1&0&0&0}, \quad 
\braket{\tau^\mu\vert\chi^0\vert\sigma^\nu}=-\bb 1i.
\ee
We know that
\bea
&&\chi^x_L\chi^x_R\ket{\sigma_L}\ket{\sigma_R}=-\ket{\tau^x_L}\ket{\tau^x_R}, 
\qquad 
\chi^x_L\chi^x_R\ket{\tau_L}\ket{\tau_R}=+\ket{\sigma^x_L}\ket{\sigma^x_R}
\\
&&\chi^y_L\chi^y_R\ket{\sigma_L}\ket{\sigma_R}=-\ket{\tau^y_L}\ket{\tau^y_R}, 
\qquad 
\chi^y_L\chi^y_R\ket{\tau_L}\ket{\tau_R}=+\ket{\sigma^y_L}\ket{\sigma^y_R}
\\
&&\chi^z_L\chi^z_R\ket{\sigma_L}\ket{\sigma_R}=-\ket{\tau^z_L}\ket{\tau^z_R}, 
\qquad 
\chi^z_L\chi^z_R\ket{\tau_L}\ket{\tau_R}=+\ket{\sigma^z_L}\ket{\sigma^z_R}.
\eea
In total there are 8x8=64 states, but the Hamiltonian is sparse. We can decouple it into a left-right diagonal sector with even parity, a left-right diagonal sector with odd parity and a remaining left-right off-diagonal sector. It turns out the diagonal elements have lower energy and remain so when $t$ becomes zero. One can show that
\be
H_{de}=\mat{\frac{\bra{\sigma_L}\bra{\sigma_R}+i\bra{\tau_L}\bra{\tau_R}}{\sqrt 2} \\ \frac{i\bra{\sigma_L}\bra{\sigma_R}+\bra{\tau_L}\bra{\tau_R}}{\sqrt 2}\\ 
\frac{\bra{\vec\sigma_L}\cdot\bra{\vec\sigma_R}+i\bra{\vec\tau_L}\cdot\bra{\vec\tau_R}}{\sqrt 6} \\ 
\frac{i\bra{\vec\sigma_L}\cdot\bra{\vec\sigma_R}+\bra{\vec\tau_L}\cdot\bra{\vec\tau_R}}{\sqrt 6}}\dg
\left\{-\frac{1}{2}\bar J\bb 1+\matc{cc|cc}{-\bar J-t/2 & 0 & \sqrt{3}t/2 & 0 \\ 0 & -\bar J+t/2 & 0 & -\sqrt{3}t/2 \\ \hline \sqrt{3}t/2 & 0& \bar J-3t/2 & 0  \\ 0 & -\sqrt{3}t/2 & 0 & \bar J+3t/2}\right\}
\mat{\frac{\bra{\sigma_L}\bra{\sigma_R}+i\bra{\tau_L}\bra{\tau_R}}{\sqrt 2} \\ \frac{i\bra{\sigma_L}\bra{\sigma_R}+\bra{\tau_L}\bra{\tau_R}}{\sqrt 2}\\ 
\frac{\bra{\vec\sigma_L}\cdot\bra{\vec\sigma_R}+i\bra{\vec\tau_L}\cdot\bra{\vec\tau_R}}{\sqrt 6} \\ 
\frac{i\bra{\vec\sigma_L}\cdot\bra{\vec\sigma_R}+\bra{\vec\tau_L}\cdot\bra{\vec\tau_R}}{\sqrt 6}}+\dots,
\ee
where the $\dots$ refers to decoupled high energy sectors. 
The low energy sector can be diagonalized by two independent orthogonal transformations
\be
U=\mat{\cos\alpha & \sin\alpha \\ -\sin\alpha & \cos\alpha}, \qquad \tan2\alpha_+=\frac{\sqrt{3}t}{2\bar J-t}, \so
E_{+,\pm}=-\frac{1}{2}\bar J-t\pm\frac{1}{2}\sqrt{(2\bar J-t)^2+3t^2},
\ee
whereas the other sector is obtained by a $t\to -t$ substitution
\be
\tan2\alpha_-=-\frac{\sqrt{3}t}{2\bar J+t}, \so
E_{-,\pm}=-\frac{1}{2}\bar J+t\pm\frac{1}{2}\sqrt{(2\bar J+t)^2+3t^2}.
\ee
The two GSs are
\bea
\ket{\psi(\alpha)}_+^e=\cos\alpha_+\ket{A}_+^e-\sin\alpha_+\ket{B}_+^e,\qquad
\ket{\psi(\alpha)}_-^e=\cos\alpha_-\ket{A}_-^e-\sin\alpha_-\ket{B}_-^e,
\eea
where we have defined
\be
\ket{A}_\pm^e=\frac{\ket{\sigma_L}\ket{\sigma_R}\pm i\ket{\tau_L}\ket{\tau_R}}{\sqrt 2}, \qqquad 
\ket{B}_\pm^e=\frac{\ket{\vec\sigma_L}\cdot\ket{\vec\sigma_R}-i\ket{\vec\tau_L}\cdot\ket{\vec\tau_R}}{\sqrt 6}.
\ee
We can do the same with the diagonal odd sector:
\be
{\cal H}_{do}=\mat{\frac{\bra{\sigma_L}\bra{\tau_R}-i\bra{\tau_L}\bra{\sigma_R}}{\sqrt 2} \\ 
\frac{\bra{\tau_L}\bra{\sigma_R}-i\bra{\sigma_L}\bra{\tau_R}}{\sqrt 2} \\ 
\frac{\bra{\vec\sigma_L}\cdot\bra{\vec \tau_R}-i\bra{\vec\tau_L}\bra{\vec\sigma_R}}{\sqrt 6}\\
\frac{\bra{\vec\tau_L}\cdot\bra{\vec \sigma_R}-i\bra{\vec\sigma_L}\bra{\vec\tau_R}}{\sqrt 6}
}\dg
\left\{-\frac{1}{2}\bar J\bb 1+\matc{cc|cc}
{-\bar J+t & 0 & \sqrt{3}t & 0 \\ 
0 & -\bar J-t & 0 & -\sqrt{3}t \\ \hline
\sqrt{3}t & 0 & \bar J-3t & 0\\
0 & -\sqrt{3}t & 0 & \bar J+3t
}\right\}
\mat{\frac{\bra{\sigma_L}\bra{\tau_R}-i\bra{\tau_L}\bra{\sigma_R}}{\sqrt 2} \\ 
\frac{\bra{\tau_L}\bra{\sigma_R}-i\bra{\sigma_L}\bra{\tau_R}}{\sqrt 2} \\ 
\frac{\bra{\vec\sigma_L}\cdot\bra{\vec \tau_R}-i\bra{\vec\tau_L}\bra{\vec\sigma_R}}{\sqrt 6}\\
\frac{\bra{\vec\tau_L}\cdot\bra{\vec \sigma_R}-i\bra{\vec\sigma_L}\bra{\vec\tau_R}}{\sqrt 6}
}+\dots
\ee
So, again we get two sectors
\bea
&&E_{-,\pm}=-\frac{1}{2}\bar J-t\pm\sqrt{(\bar J-2t)^2+3t^2},\qquad \tan2\alpha_-=\frac{\sqrt{3}t}{\bar J-2t},\\
&&E_{+,\pm}=-\frac{1}{2}\bar J+t\pm\sqrt{(\bar J+2t)^2+3t^2},\qquad \tan2\alpha_+=-\frac{\sqrt{3}t}{\bar J+2t},
\eea
with the GS wavefunctions
\be
\ket{\psi(\alpha)}_-^o=\cos\alpha_-\ket{X}_-^o-\sin\alpha_-\ket{Y}_-^o, \andd
\ket{\psi(\alpha)}_+^o=\cos\alpha_+\ket{X}_+^o-\sin\alpha_+\ket{Y}_+^o,
\ee
where we have defined
\be
\ket{X}_\pm^o=\frac{\ket{\sigma_L}\ket{\tau_R}\pm i\ket{\tau_L}\ket{\sigma_R}}{\sqrt 2}, \qqquad 
\ket{Y}_\pm^o=\frac{\ket{\vec\sigma_L}\cdot\ket{\vec\tau_R}-i\ket{\vec\tau_L}\cdot\ket{\vec\sigma_R}}{\sqrt 6}.
\ee
\subsubsection{Entangled states of Kondo MZMs}
On the other hand, suppose before turning on $t$ we have initialized the two MZMs using the following Hamiltonain.
\be
H_{\rm initialization}=im\gamma_L\gamma_R.
\ee
For $m>0$ the two GSs ($E=-m/2$) are
\be
\ket{\phi}_+^e=\frac{\ket{\sigma_L}\ket{\sigma_R}+i\ket{\tau_L}\ket{\tau_R}}{\sqrt{2}},\qqquad
\ket{\phi}_+^o=\frac{\ket{\sigma_L}\ket{\tau_R}+i\ket{\tau_L}\ket{\sigma_R}}{\sqrt{2}}, 
\ee
and the two excited states ($E=m/2$) are
\be
\ket{\phi}_-^e=\frac{\ket{\sigma_L}\ket{\sigma_R}-i\ket{\tau_L}\ket{\tau_R}}{\sqrt{2}},\qqquad
\ket{\phi}_-^o=\frac{\ket{\sigma_L}\ket{\tau_R}-i\ket{\tau_L}\ket{\sigma_R}}{\sqrt{2}}.
\ee
whereas for $m<0$ the two sets of states flip.
\subsubsection{Fusion: Mapping Kondo MZMs' states to spin states}
Since nothing modifies the fermion parity, we can assume to be in one sector, say the even fermion parity sector. Then the sign of $m$ determines the GS of $H_{\rm init.}$. Then, turning $t$ and ramping it up, the angles evolve as
\be
2\alpha_+:\quad 0\,\,(\to \pi/2)\,\to 2\pi/3, \andd 2\alpha_-:\quad 0\to -\pi/3.
\ee
Therefore, we would unambigiously end up at the following states
\bea
&&\ket{\psi(\alpha)}_+^e=\qquad \ket{X}_-^e\Big\vert_{\alpha=0}
\quad\to\quad \frac{1}{2}\ket{X}_+^e-\frac{\sqrt{3}}{2}\ket{Y}_+^e\Big\vert_{\alpha=\pi/3}, \\
&&\ket{\psi(\alpha)}_-^e=\qquad \ket{X}_-^e\Big\vert_{\alpha=0}
\quad\to\quad \frac{\sqrt 3}{2}\ket{X}_-^e+\frac{1}{2}\ket{Y}_-^e\Big\vert_{\alpha=-\pi/6}.
\eea
In the space of $\ket{X}_\pm$ and $\ket{Y}_{\pm}$ the $\vec S_1\cdot\vec S_2$ is (independently of the $\pm$ sectors)
\be
\mat{\ket{X}_\pm\\ \ket{Y}_\pm}\dg\vec S_1\cdot\vec S_2\mat{\ket{X}_\pm\\ \ket{Y}_+}=\frac{1}{2}\mat{0 & \sqrt{3}/2 \\ \sqrt{3}/2 & -1}.
\ee
Now, assuming $J/t\to 0$ the spins are decoupled and we can measure
\be
{}_+^e\sbraket{\psi\vert \vec S_1\cdot\vec S_2\vert\psi}_+^e:\quad 0\to -\frac{3}{4}, \andd
{}_-^e\sbraket{\psi\vert \vec S_1\cdot\vec S_2\vert\psi}_-^e:\quad 0 \to \frac{1}{4}.
\ee
\subsection{Braiding -- Non-interacting particles}
\begin{figure}
\includegraphics[width=0.8\textwidth]{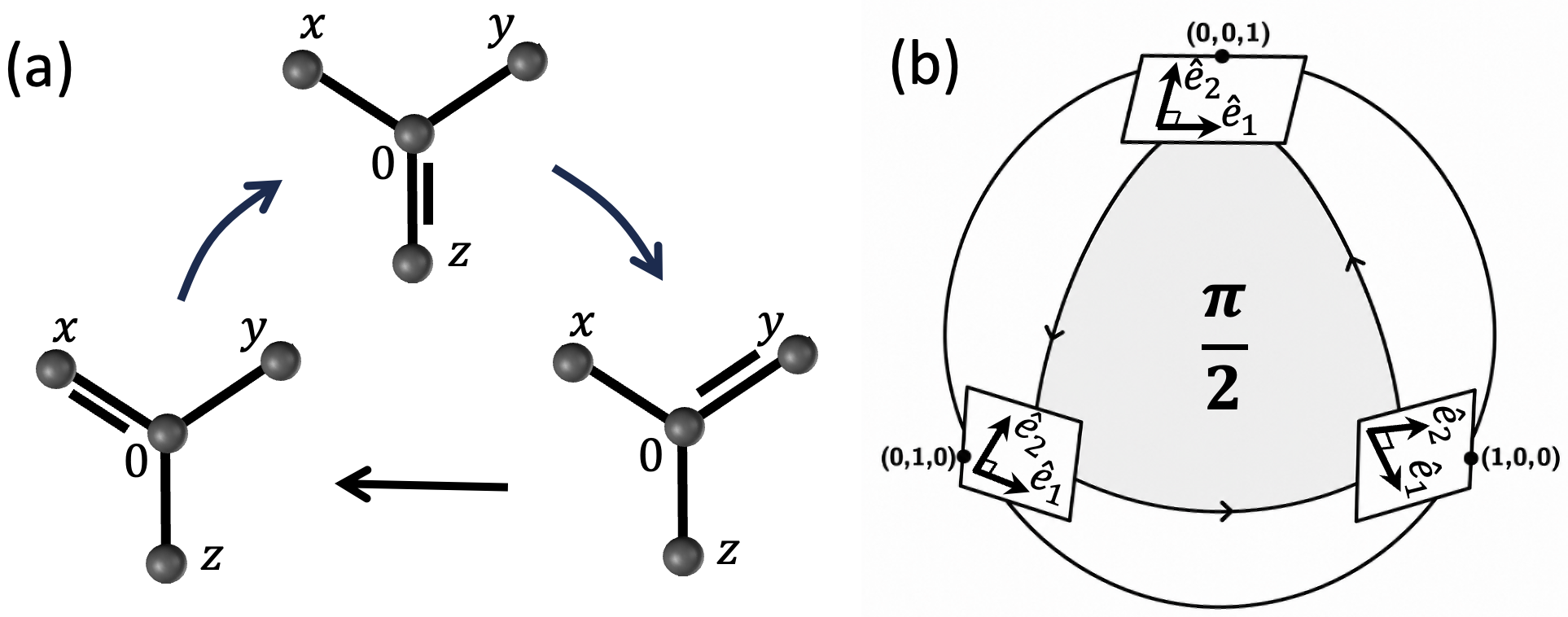}
\caption{\raggedright\small (a) General setting for braiding non-interacting fermions. (b) Parallel transport of the tangent plane unit vectors along a loop on the unit sphere.}\label{FigS1}
\end{figure}

Braiding requires a minimal two-dimensional geometry, naturally realized by the Y-junctions shown in Fig.\,4 of the main text. In all cases, we initialize two detached Kondo MZMs on the $x$ and $y$ legs and perform the cyclic transfer sequence
$y\to z$, $x\to y$, and $z\to x$, thereby exchanging the two Kondo anyons [Fig.\,\ref{FigS1}].

We consider a $c$-centered $Y$ junction, composed of four fermions, described by the Hamiltonian
\be
H(t)=c\dg_0[v_xc_x+v_yc_y+v_zc_z]+h.c.,
\ee
with time-dependent tunneling amplitudes $v_\mu(t)$. We are using $\vec v$ to refer to the hybridization vector of the central $c_0$ mode to differentiate tunneling amplitudes from the time $t$ in this section. Without loss of generality we take the $v$-s to be real. At any given time, only one linear combination of fermions, the bright mode $b(t)$, is coupled to the central $c_0$ fermion. $H(t)\propto [c\dg_0 b(t)+h.c.]$. This leaves two dark (gapless) fermions $d_1$ and $d_2$, that span the low-energy Hilbert space. 
\subsubsection{Holonomy}
A general construction is to pick a unit vector $\hat n$ that is never parallel to $\hat v$. Then the two unit vectors $\hat e_1$ and $\hat e_2$ provide a natural local frame for the dark sector (or the local chart of dark manifold),
\be
\hat v=\frac{(v_x,v_y,v_z)}{\sqrt{v_x^2+v_y^2+v_z^2}}, \qqquad \hat e_1=\frac{\hat n\times\hat v}{\abs{\hat n\times\hat v}},\qqquad
\hat e_2=\frac{\hat n-(\hat n\cdot\hat v)\hat v}{\abs{\hat n\times\hat v}}.
\ee
These satisfy $\hat e_1\cdot\hat v=\hat e_2\cdot\hat v=\hat e_1\cdot\hat e_2=0$, which guarantees  that the corresponding fermions remain orthogonal
\be
b=\hat v\cdot\vec c, \quad d_1=\hat e_1\cdot\vec c, \quad d_2=\hat e_2\cdot\vec c \so 
\{d_1,d_2\dg\}=\{d_1,b\dg\}=\{d_2,b\dg\}=0.
\ee
In the tangent plane spanned by $(\hat e_1,\hat e_2)$, the parallel transport connection is

\be
A_{ij}=\hat e_i\cdot d\hat e_j=\mat{0 & -\omega \\ \omega & 0}, \qquad \omega=\frac{(\hat n\cdot\hat v)(\hat n\times \hat v)\cdot d\hat v}{\abs{\hat n\times\hat v}^2}=\frac{a}{1-a^2}\eta, \qquad a\equiv\hat n\cdot\hat v, \quad \eta\equiv\hat n\cdot(\hat v\wedge d\hat v).
\ee
To use Stokes theorem, we need to compute $d\omega$
\be
d\omega=\frac{1+a^2}{(1-a^2)^2}da\wedge\eta+\frac{a}{1-a^2}d\eta, \qquad d\eta=\hat n\cdot (d\hat v\wedge d\hat v)=2a\Omega_{S^2}, \qquad da\wedge\eta=-(1-a^2)\Omega_{S^2},
\ee

\noindent written in terms of the standard area 2-form on the unit sphere:
\be
\Omega_{S^2}=\frac{1}{2}\hat v\cdot (d\hat v\wedge d\hat v)=\frac{1}{2}\eps_{ijk}\hat v_i\cdot (d\hat v_j\wedge d\hat v_k)
\ee
To verify this, we remind the reader that in spherical coordinates $d\hat r=dr\hat r+rd\theta\hat\theta+r\sin\theta d\varphi\hat\varphi$. Summing up, we find $d\omega=-\Omega_{S^2}$, 
leading to the enclosed integral
\be
\oint_{\partial\cal A}{\vec A\cdot\vec d\ell}=\mat{0 & -1 \\ 1 & 0}\int_{\cal A}d\omega=-\frac{\pi}{2}\mat{0 & -1 \\ 1 & 0},
\ee
and the holonomy
\be
W=\exp[-\oint_{\partial\cal A}{\vec A\cdot\vec d\ell}]=\exp[-i\frac{\pi}{2}\sigma^y]=-i\sigma^y.
\ee
\subsubsection{Evolution operator}
The previous holonomy naturally emerges in the evolution operator
\be
\hat U=\exp\Big[-i\int_0^T dt H(t)\Big].
\ee
Since $H(t)=c\dg{\cal H}(t) c$, is non-interacting, fermion evolution is determined by the single-particle evolution operator
\be
U={\cal P}\exp\Big[-i\int_0^T dt {\cal H}(t)\Big] \So \hat U\dg c \hat U=U\vec c.
\ee
The trotter discretization of $U$ gives
\be
U=\prod_{k=0}^{N-1}e^{-i{\cal H}(t_k)\Delta t}=\prod_{k=0}^{N-1}{\cal U}_k e^{-i\tilde {\cal H}(t_k)\Delta t}{\cal U}_k^T, \qquad {\cal U}\equiv\matc{ccc}{\hat v & \hat e_1 & \hat e_2},
\ee
where we have used ${\cal U}_k\tilde{\cal H}(t_k){\cal U}_k^T$ to diagonalize ${\cal H}_k$.  Using the the fact that ${\cal U}^T_k{\cal U}_{k-1}=\bb 1-\Delta t \bb A(t_k)+O(\Delta t^2)$,
\be
U=\prod_k e^{-i\tilde {\cal H}(t_k)\Delta t}[\bb 1-\Delta t \bb A(t_k)].
\ee
But these matrices do not commute. To simplify, we project to the dark space, where $\tilde{\cal H}=0$. Then,
\be
PUP=\prod_k [\bb 1-\Delta t \bb A(t_k)]=\exp\Big[-\int_0^T dt A(t)\Big], \qquad A=P\bb A P,
\ee
which is the holonomy $W$ we computed before. Therefore, we find $U=W=-i\sigma^y$ or
\be
\hat U\dg\mat{c_x \\ c_y}\hat U=U\mat{c_x\\ c_y}=\mat{-c_y \\ c_x}.
\ee
This discussion was quite general, not even involving the statistics of non-interacting particles. As such, it is applicable not only to complex (Dirac) spinless fermions, but also to spinful fermions, Majorana fermions or even bosons.
\subsubsection{A smooth connection}
A convenient choice, allowing to chart the whole evolution loop is $\hat n={(1,1,1)}/{\sqrt3}$ which leads to
\be
\hat e_1\sim\frac{(v_z-v_y,v_x-v_z,v_y-v_x)}{\sqrt{(v_z-v_y)^2+(v_x-v_z)^2+(v_y-v_x)^2}}, \quad 
\hat e_2\sim \frac{[v_y^2+v_z^2-v_x(v_y+v_z)],[v_x^2+v_z^2-v_y(v_x+v_z)],[v_x^2+v_y^2-v_z(v_x+v_y)]}{\sqrt{\dots}}.
\ee
For the first leg, $y\to z$, we have $t_x=0$
\be
\hat e_1\sim\frac{(v_z-v_y,-v_z,v_y)}{\sqrt{(v_z-v_y)^2+v_z^2+v_y^2}}:\quad  \frac{(1,-1,0)}{\sqrt{2}}\to\frac{(-1,0,1)}{\sqrt 2}
\qquad 
\hat e_2\sim \frac{[v_y^2+v_z^2],v_z[v_z-v_y],v_y[v_y-v_z]}{\sqrt{(v_y^2+v_z^2)[(v_z^2+v_y^2)+(v_y-v_z)^2]}}:\quad\frac{(1,1,0)}{\sqrt 2}\to \frac{(1,0,1)}{\sqrt 2}.
\ee
For the second leg, $x\to y$, we have $v_z=0$
\be
\hat e_1\sim\frac{(-v_y,v_x,v_y-v_x)}{\sqrt{v_y^2+v_x^2+(v_y-v_x)^2}}:\quad \frac{(-1,0,1)}{\sqrt 2}\to\frac{(0,1,-1)}{\sqrt2} 
\qquad 
\hat e_2\sim \frac{v_y[v_y-v_x],v_x[v_x-v_y],[v_x^2+v_y^2]}{\sqrt{(v_x^2+v_y^2)[v_x^2+v_y^2+(v_x-v_y)^2]}}:\quad\frac{(1,0,1)}{\sqrt 2}\to\frac{(0,1,1)}{\sqrt 2}.
\ee
And for the third leg, $z\to x$, we have $v_y=0$
\be
\hat e_1\sim\frac{(v_z,v_x-v_z,-v_x)}{\sqrt{v_z^2+(v_x-v_z)^2+v_x^2}}: \quad 
\frac{(0,1,-1)}{\sqrt2}\to\frac{(1,-1,0)}{\sqrt2},\qquad
\hat e_2\sim \frac{v_z[v_z-v_x],[v_x^2+v_z^2],v_x[v_x-v_z]}{\sqrt{(v_x^2+v_z^2)[v_x^2+v_z^2+(v_z-v_x)^2]}}:\quad
\frac{(0,1,1)}{\sqrt 2}\to\frac{(1,1,0)}{\sqrt 2},
\ee
which is indeed continuous. For this choice, the line-integral is distributed evenly through the loop.

\subsubsection{An almost trivial connection}
Another special case, more useful for the Kondo interaction, is to choose $\hat n=-\hat x$. Then,
\be
b=\hat v\cdot\vec c=\frac{v_xc_x+v_yc_y+v_zc_z}{\sqrt{v_x^2+v_y^2+v_z^2}}, \qquad  d_{1} \equiv\hat e_1\cdot\vec c=\frac{v_zc_y-v_yc_z}{\sqrt{v_y^2+v_z^2}}, \qquad d_2\equiv \hat e_2\cdot\vec c=\frac{(v_y^2+v_z^2)c_x-v_xv_yc_y-v_xv_zc_z}{\sqrt{v_y^2+v_z^2}\sqrt{v_x^2+v_y^2+v_z^2}}.
\ee
The issue with this choice of local frame is that for $v_y=v_z=0$, i.e. after the $x\to y$ transfer but before the $z\to x$ transfer, both $d_1$ and $d_2$ fermions are not well-defined. This can be fixed by patching the two charts together. We have
\be
d_{yz}=\frac{v_zc_y-v_yc_z}{\sqrt{v_y^2+v_z^2}}\Big\vert_{v_x=0}=d_1,\qquad 
d_{xy}=\frac{v_yc_x-v_xc_y}{\sqrt{v_y^2+v_x^2}}\Big\vert_{v_z=0}=d_2,\qquad 
d_{zx}=\frac{v_xc_z-v_zc_x}{\sqrt{v_x^2+v_z^2}}\Big\vert_{v_y=0}=-d_2.
\ee
These states evolve as
\be
\mat{c_x \\ c_y \\ c_z}:\qquad
 \mat{d_2\\ d_1 \\ b}\xrightarrow{y\to z} \mat{d_2\\ b \\ -d_1},\quad 
\mat{d_2 \\ b \\ -d_1}\xrightarrow{x\to y} \mat{b \\ -d_2 \\ -d_1},\quad 
\xRightarrow{S_{xy\to zx}}
\quad
\mat{b \\ d_1 \\ -d_2}\xrightarrow{z\to x} \mat{d_2 \\ d_1 \\ b}.
\ee
To match the basis at the end of the second step with that of  the beginning of the third step, an $S$-matrix is needed:
\be
\mat{b \\ d_1 \\ -d_2}=S_{xy\to zx}\mat{b \\ -d_2 \\ -d_1},\qquad
\mat{\hat v \\ \hat e_1 \\ -\hat e_2}=S_{xy\to zx}\mat{\hat v \\ -\hat e_2 \\ -\hat e_1},\qquad
S_{xy\to zx}=\matc{c|cc}{1 \\ \hline & & -1 \\ & 1}.
\ee
Along each branch 
\be
\mat{c_y\\ c_z}\propto\mat{v_y & v_z \\ v_z & -v_y}^{-1}\mat{b\\ d_1},\quad
\mat{c_x\\ c_y}\propto\mat{v_x & v_y \\ v_y & -v_x}^{-1}\mat{b\\ d_2},\quad
\mat{c_z\\ c_x}\propto\mat{v_z & v_x \\ v_x & -v_z}^{-1}\mat{b\\ -d_2},
\ee
or in the normalized form
\be
\mat{d_1 \\b}=O(\theta_1)\mat{c_y\\ c_z},\quad
\mat{d_2 \\b}=O(\theta_2)\mat{c_x\\ c_y},\quad
\mat{-d_2 \\b}=O(\theta_3)\mat{c_z\\ c_x}, \qquad O(\theta_i)=\mat{ \cos\theta_i & -\sin\theta_i \\\sin\theta_i & \cos\theta_i}.
\ee
In this case, the connection throughout each section of the path is:
\be
\bb A=O^T\partial_\theta O=\mat{ \cos\theta_i & \sin\theta_i \\-\sin\theta_i & \cos\theta_i}\mat{ \sin\theta_i & \cos\theta_i \\-\cos\theta_i & \sin\theta_i}d\theta_i=\mat{0 & -1 \\ 1 & 0}d\theta_i \so A=P_d \bb A P_d=0,
\ee
which vanishes because $\bb A$ always involves the bright mode. The whole holonomy comes from the chart-sewing part 
\be
W_{dark}=PWP=S_{xy\to zx}=\mat{& -1 \\ 1 &}=-i\sigma^y,
\ee
which is the same as before. This flat connection is more useful for the interacting case, treated below.
\subsection{Braiding -- Kondo MZMs}
Neglecting the gapped $b$-fermion, the Kondo interaction is
\be
H=\vec{\cal J}[d_1]\cdot[\cos^2_1\theta S_y+\sin^2_1\theta S_z]+\vec{\cal J}[d_2]\cdot[\cos^2_2\theta S_x+\sin^2_2\theta S_y]+\vec{\cal J}[d_2]\cdot[\cos^2_3\theta S_z+\sin^2_3\theta S_x].
\ee
With the GS
\bea
&&\kettt{\Ua}^{o/e}\equiv\cos\alpha \Big[\sqrt{2/3}\ket{1,1}\ket{\da/0}-\sqrt{1/3}\ket{1,0}\ket{\ua/2}\Big]-\sin\alpha\ket{S}\ket{\ua/2},
\\
&&\kettt{\Da}^{o/e}\equiv\cos\alpha\Big[\sqrt{1/3}\ket{1,0}\ket{\da/0}-\sqrt{2/3}\ket{1,-1}\ket{\ua/2}\Big]-\sin\alpha\ket{S}\ket{\da/0},
\eea
with the angle $\alpha$ related to $\theta$ according to
 \be
 \tan2\alpha=\frac{2{V}}{\bar J}, \qquad V=-\sqrt{3}\delta J/4 \so  \tan2\alpha=-\sqrt{3}\cos2\theta.
 \ee
As $\theta:0\to \pi/2$ we have $\alpha:-\pi/6\to \pi/6$. It is helpful to express the GSs $\ket{\sigma_{ij}}$ and $\ket{\tau_{ij}}$ in terms of spin-$i$ and fermion-$j$ which have formed a singlet. Then, the eight-fold degenerate GS is
\be
\ket{\Psi}=\mat{\ket{\sigma_{xx}}\ket{\sigma_{yy}} \\ \ket{\sigma_{xx}}\ket{\tau_{yy}} \\ \ket{\tau_{xx}}\ket{\sigma_{yy}} \\ \ket{\tau_{xx}}\ket{\tau_{yy}}}\otimes\Big\{\ket{\Ua,\Da}_z\Big\}=\mat{\ket{\sigma_{x2}}\ket{\sigma_{y1}} \\ \ket{\sigma_{x2}}\ket{\tau_{y1}} \\ \ket{\tau_{x2}}\ket{\sigma_{y1}} \\ \ket{\tau_{x2}}\ket{\tau_{y1}}}\otimes\Big\{\ket{\Ua,\Da}_z\Big\}.
\ee
In the space of $\ket{\Psi}$, the Kondo MZMs are represented as $\hat\gamma_{x,y}=\ket{\Psi}\gamma_{x,y}\bra{\Psi}$ where
\be
\gamma_x=\frac{1}{\sqrt 2}\matc{cc|cc}{ & -1 & \\ -1 &  & \\ \hline && & 1 \\ &&1}=-\frac{\sigma^x\tau^z}{\sqrt 2},\qquad
\qquad 
\gamma_y=\frac{1}{\sqrt 2}\matc{cc|cc}{ & & 1 \\  &  & & 1\\ \hline 1&& & \\ &1&&}=-\frac{\tau^x}{\sqrt 2},
\ee
which obey $\gamma_x^2=\gamma_y^2=\bb1/\sqrt{2}$ and $\gamma_x\gamma_y+\gamma_x\gamma_y=0$. The unitary transformations of the braiding operation lead to
\bea
&&\mat{\ket{\sigma_{x2}}\ket{\sigma_{y1}} \\ \ket{\sigma_{x2}}\ket{\tau_{y1}} \\ \ket{\tau_{x2}}\ket{\sigma_{y1}} \\ \ket{\tau_{x2}}\ket{\tau_{y1}}}\otimes\Big\{\ket{\Ua,\Da}_z\Big\} 
\xrightarrow{y\to z}
\mat{\ket{\sigma_{x2}}\ket{\sigma_{z1}} \\ \ket{\sigma_{x2}}\ket{\tau_{z1}} \\ \ket{\tau_{x2}}\ket{\sigma_{z1}} \\ \ket{\tau_{x2}}\ket{\tau_{z1}}}\otimes\Big\{\ket{\Ua,\Da}_y\Big\}
\xrightarrow{x\to y}
\mat{\ket{\sigma_{y2}}\ket{\sigma_{z1}} \\ \ket{\sigma_{y2}}\ket{\tau_{z1}} \\ \ket{\tau_{y2}}\ket{\sigma_{z1}} \\ \ket{\tau_{y2}}\ket{\tau_{z1}}}\otimes\Big\{\ket{\Ua,\Da}_x\Big\}\\
&&\hspace{1cm}
\xRightarrow{S_{xy\to zx}}
\mat{-\ket{\sigma_{y1}}\ket{\sigma_{z2}} \\ \ket{\sigma_{y1}}\ket{\tau_{z2}} \\ -\ket{\tau_{y1}}\ket{\sigma_{z2}} \\ \ket{\tau_{y1}}\ket{\tau_{z2}}}\otimes\Big\{\ket{\Ua,\Da}_x\Big\}
\xrightarrow{z\to x}
\mat{-\ket{\sigma_{y1}}\ket{\sigma_{x2}} \\ \ket{\sigma_{y1}}\ket{\tau_{x2}} \\ -\ket{\tau_{y1}}\ket{\sigma_{x2}} \\ \ket{\tau_{y1}}\ket{\tau_{x2}}}\otimes\Big\{\ket{\Ua,\Da}_z\Big\}
=
\mat{\ket{\sigma_{x2}}\ket{\sigma_{y1}} \\ \ket{\tau_{x2}}\ket{\sigma_{y1}} \\ -\ket{\sigma_{x2}}\ket{\tau_{y1}} \\ \ket{\tau_{x2}}\ket{\tau_{y1}}}\otimes\Big\{\ket{\Ua,\Da}_z\Big\}.\nonumber
\eea
Therefore, the unitary transformation $\hat U$ acts on the $\ket{\Psi}$ space as
\be
\ket{\Psi}\to U\ket{\Psi}, \qquad U=\matc{cc|cc}{1 & & \\ & & 1 \\ \hline & -1 & \\ &&&1} \So
U\dg\gamma_x U=\gamma_y, \qquad U\dg \gamma_y U=-\gamma_x.
\ee
\subsection{Numerical Results}
\begin{figure}[h!]
\includegraphics[width=0.51\linewidth]{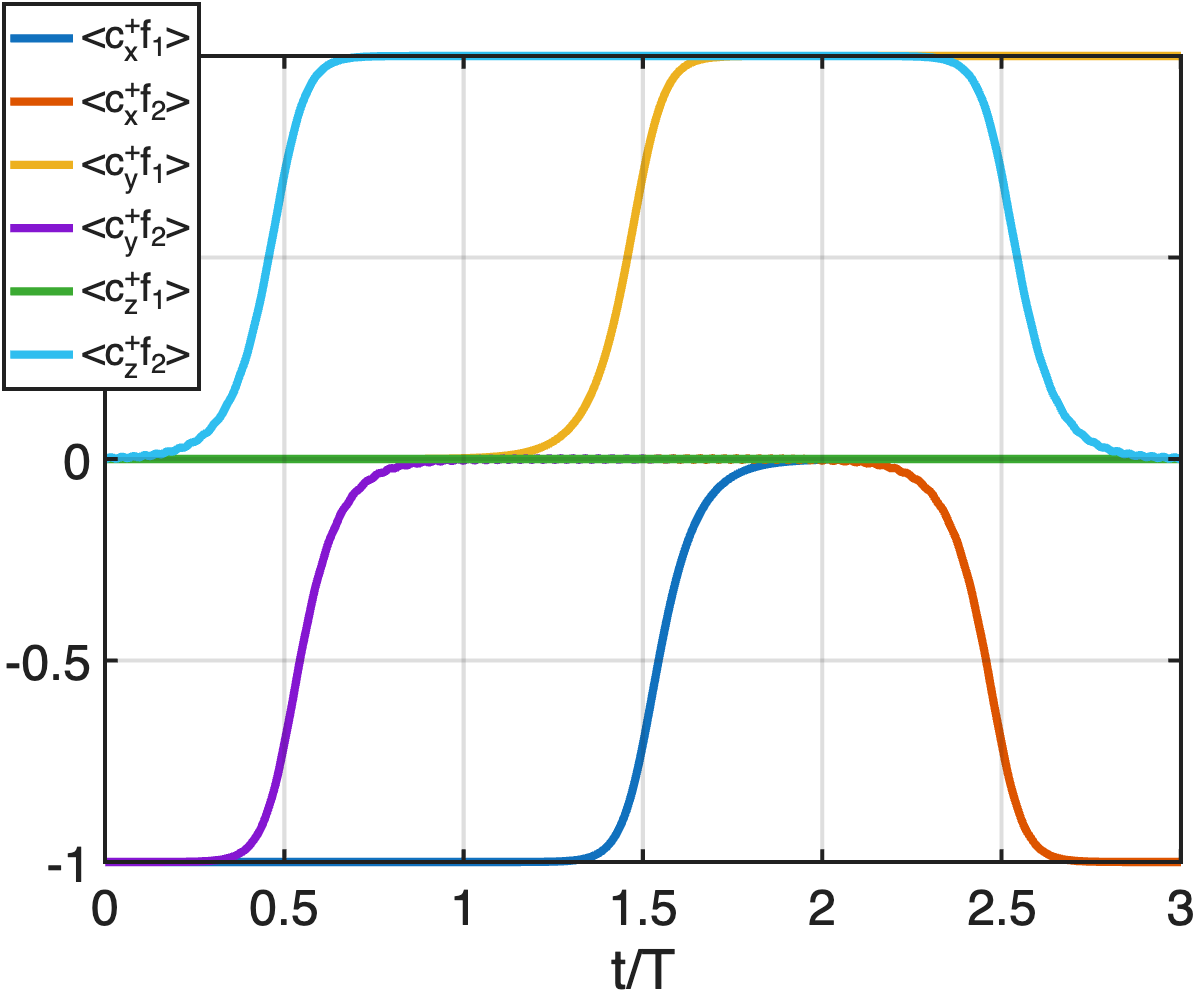}
\includegraphics[width=0.45\linewidth]{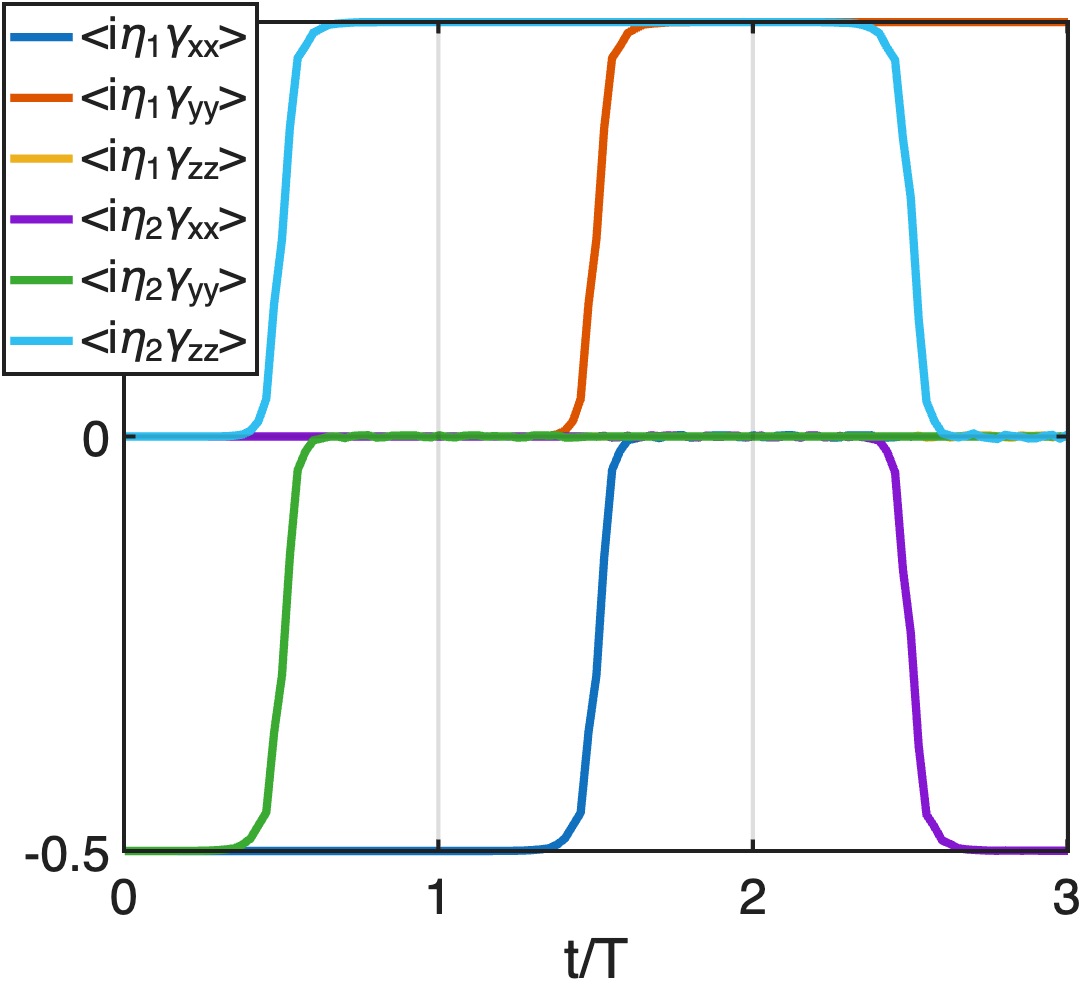}
\caption{\small Braiding in a $c$-centered Y junction; (left) non-interacting fermions, (right) Kondo MZMs}\label{FigS2}
\end{figure}

Fig.\,\pref{FigS2} shows the effect of braiding on both non-interacting complex (Dirac) fermions and Kondo MZMs in a $c$-centered Y junction.

\section{\sc\large Quantum Gates}
For completeness, in this section we list a number of known results from the literature \cite{Barenco1995,Boykin1999,Georgiev2006,Bravyi2006,Nielsen2012,Beenakker2020,Zhang2024d}. 
\subsection{Braiding-based Hadamard and CNOT gates}
\begin{figure}[h!]
\includegraphics[width=0.9\linewidth]{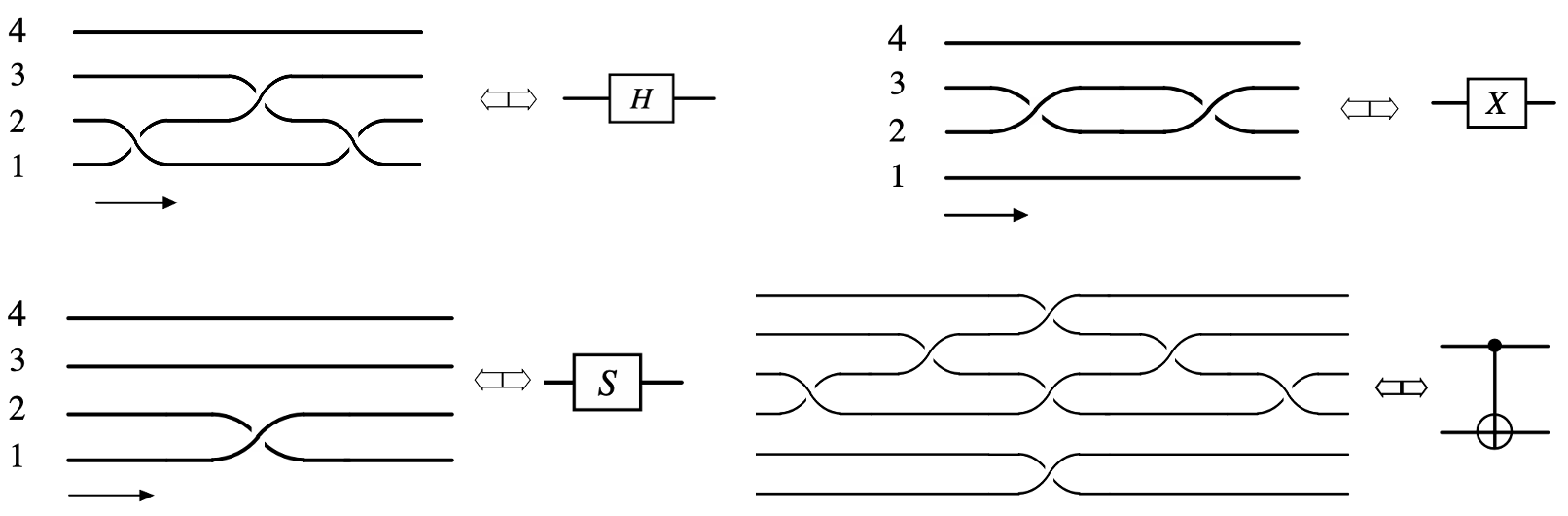}
\caption{\small Braiding approach to single- and two-qubit gates \cite{Georgiev2006}.}\label{FigS3}
\end{figure}

The braiding operation $R_{ij}$ can be used \cite{Georgiev2006} to realize single-qubit operations
\be
H\cong R\dn_{13}R_{12}^2=R_{12}^{-1}R_{23}\dn R_{12}^{-1}=\frac{1}{\sqrt 2}\mat{1 & 1 \\ 1 & -1}, \qquad X=R_{23}^2=\mat{0 & 1 \\ 1 &0},\qquad S=R_{12}=R_{34}=\mat{1 & 0 \\ 0 & i},
\ee
as well as a two-qubit CNOT operation
\be
{\rm CNOT}=R_{34}^{-1}R_{45}\dn R_{12}\dn R_{56}\dn R_{45}\dn R_{34}^{-1}=\matc{cc|cc}{1 & 0 & 0 & 0\\ 0 & 1 & 0 & 0 \\ \hline 0 & 0& 0 & 1 \\ 0 &0 & 1 & 0}.
\ee
These gates all rely on nearest neighbor braidings, and depicted in Fig.\,\pref{FigS3}.

\subsection{Measurement-induced CNOT gate}
The two-qubit CNOT gate can also be achieved through parity measurement \cite{Beenakker2004,Zilberberg2008,Beenakker2020}. Let
\be
\ket{\psi}_c = a\ket{0}_c+b\ket{1}_c,
\qquad
\ket{\phi}_t=\alpha\ket{0}_t+\beta\ket{1}_t,
\qquad
\ket{+}_a=\frac{\ket{0}_a+\ket{1}_a}{\sqrt2},
\ee
where the ancillary state $\ket{+}_a=H\ket{0}_a$ can be obtained from a Hadamard gate. The initial state is
\be
\ket{\Psi_i}
=
\ket{\psi}_c\otimes\ket{+}_a \otimes  \ket{\phi}_t.
\ee
We first (projectively) measure the joint parity $Z_cZ_a$. The resulting state depends on the outcome $m_{ZZ}$
\be
\ket{\Psi_i}\xrightarrow{Z_cZ_a}\ket{\Psi_1}=
\left\{
\matl{
[a\ket{00}_{ca}+b\ket{11}_{ca}]\otimes\ket{\phi}_t\qquad  & m_{ZZ}=+1\\ 
\left[a\ket{01}_{ca}+b\ket{10}_{ca}\right]
\otimes\ket{\phi}_t & m_{ZZ}=-1
}
\right.
.
\ee
Next, we expand the target state in the $X$ basis,
\begin{equation}
|\phi\rangle_t=\gamma|+\rangle_t+\delta|-\rangle_t,
\qquad
\gamma=\frac{\alpha+\beta}{\sqrt2},
\qquad
\delta=\frac{\alpha-\beta}{\sqrt2}.
\end{equation}
Continuing with the branch $m_{ZZ}=+1$, measuring $X_aX_t$ on $\ket{\Psi_1^+}$ gives:
\be
\ket{\Psi_1^+}\xrightarrow{X_aX_c}\ket{\Psi_2^+}\propto\left\{
\matl{
\gamma(a\ket{0}+b\ket{1})_c \ket{++}_{at}+\delta(a\ket{0}-b\ket{1})_c\ket{--}_{at}\qquad  & m_{ZZ}=+1,\quad m_{XX}=+1\\ 
\delta(a\ket{0}+b\ket{1})_c \ket{+-}_{at}+\gamma(a\ket{0}-b\ket{1})_c \ket{-+}_{at} & m_{ZZ}=-1,\quad m_{XX}=-1
}
\right.
,
\ee
which again depends on the outcome $m_{XX}=+1$ and $m_{XX}=-1$. Finally, we measure the ancilla in the $Z$ basis.

\begin{figure}[h!]
\includegraphics[width=0.5\linewidth]{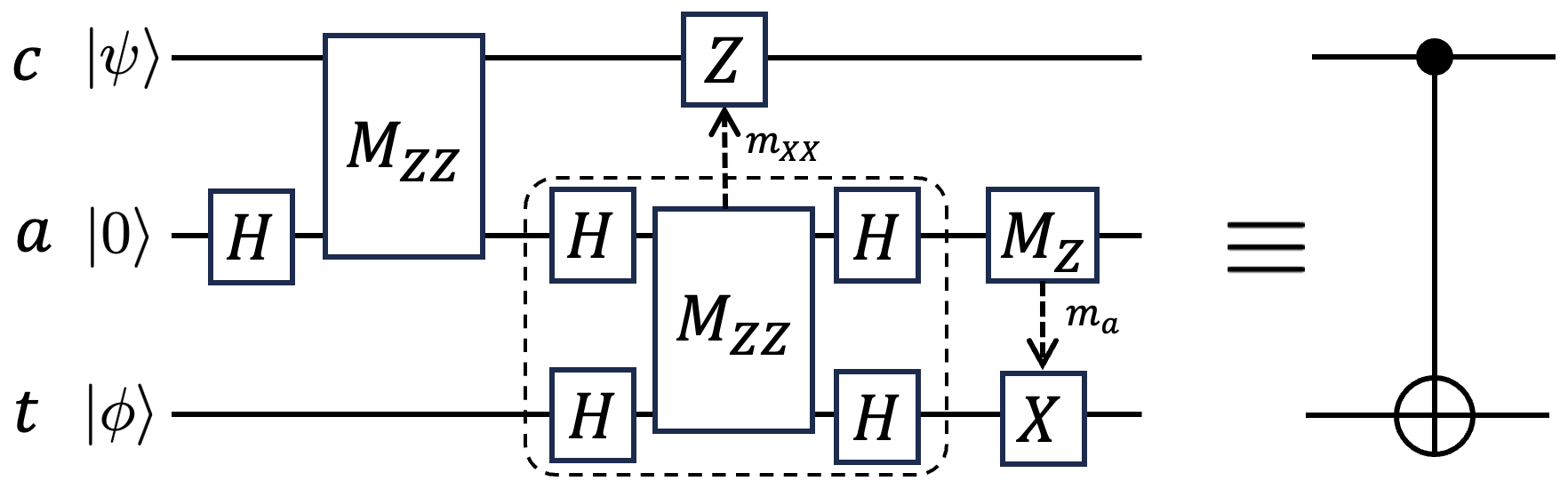}
\caption{\small Measurement based CNOT gate using an additional ancillary qubit \cite{Beenakker2004,Zilberberg2008,Beenakker2020}.}\label{FigS4}
\end{figure}

For the branch $(m_{ZZ},m_{XX})=(+1,+1)$, if $m_a=0$,
\bea
\ket{\Psi_2^{++}}\xrightarrow{Z_a}
&&\ket{\Psi_f^{++;0}}\propto
\gamma(a\ket{0}+b\ket{1})\ket{+}_t+\delta(a\ket{0}-b\ket{1})\ket{-}_t  \\
&&\hspace{2cm}=
a\ket{0}\big(\gamma\ket{+}+\delta\ket{0}\big)+
b\ket{1}\big(\gamma\ket{+}-\ket{-}\big) \\
&&\hspace{3cm}=a\ket{0}\ket{\phi}_t+b\ket{1} X_t\ket{\phi}_t =
{\rm CNOT}_{c\to t}\big(\ket{\psi}_c\otimes\ket{\phi}_t\big).
\eea
On the other hand, if $m_a=1$,
\bea
\ket{\Psi_2^{++}}\xrightarrow{Z_a}\ket{\Psi_f^{+,+;1}}
&&\propto
\gamma(a\ket{0}+b\ket{1})\ket{+}_t -\delta(a\ket{0}-b\ket{1})\ket{-}_t \\
&&=
a\ket{0} X_t\ket{\phi}_t+b\ket{1}\ket{\phi}_t = X_t\,\mathrm{CNOT}_{c\to t}\big(\ket{\psi}_c\otimes\ket{\phi}_t\big).
\eea
Thus, in this branch the only byproduct is $X_t^{m_a}$. Proceeding similarly for all branches, one finds the general result
\begin{equation}
|\Psi_{\rm out}\rangle
=
X_t^{\,m_a}\, Z_c^{\,\frac{1-m_{XX}}{2}}\,
\mathrm{CNOT}_{c\to t}
\big(|\psi\rangle_c\otimes|\phi\rangle_t\big),
\end{equation}
up to conventions for labeling the outcomes $m_{ZZ},m_{XX},m_a=\pm1$ where the powers are understood to be modulo $2$. For MZMs, the ZZ measurement is a parity measurement, whereas the XX measurement can be related to a ZZ measurement by additional Hadamard dressing; $X=HZH$. Fig.\,\pref{FigS4} summarizes this protocol

\ew

\end{document}